\documentstyle[aps,eqsecnum,epsf]{revtex}
\newcommand{\lee}{\mbox{
{${\cal L}$}\hspace*{-8pt}\raisebox{1pt}{$-$}}}

\begin{document}

\rightline{\large\baselineskip20pt\rm\vbox to20pt{
\baselineskip14pt
\hbox{OU-TAP 99}
\hbox{gr-qc/9904076}
\vspace{1mm}
\hbox{24 Sep 1999}\vss}}%
\vskip8mm
\begin{center}{\large \bf
Generalization of the model of Hawking radiation\\
with modified high frequency dispersion relation 
} 
\end{center}
\vspace*{4mm}
\centerline{\large Yoshiaki Himemoto$^{a}$\footnote
      {E-mail:himemoto@vega.ess.sci.osaka-u.ac.jp}
 and Takahiro Tanaka$^{a,b}$\footnote{E-mail:tama@vega.ess.sci.osaka-u.ac.jp}
} 
\vspace*{4mm}
\centerline{${}^a$\em Department of Earth and Space Science, Graduate 
School of Science}
\centerline{\em Osaka University, Toyonaka 560-0043, Japan} 
\centerline{${}^b$\em IFAE, Departament de F{\'\i}sica, 
  Universitat Aut{\`o}noma de Barcelona, 08193 Bellaterra, Spain}

\begin{abstract}
The Hawking radiation is one of the most interesting phenomena 
predicted by the theory of quantum field in curved space. 
The origin of Hawking radiation is closely 
related to the fact that 
a particle which marginally 
escapes from collapsing into a black hole 
is observed at the future infinity with  
infinitely large redshift. 
In other words, 
such a particle had a very high frequency 
when it was near the event horizon. 
Motivated by the possibility that 
the property of Hawking radiation may be altered  
by some unknowned physics which may exist beyond some critical scale, 
Unruh proposed a model which has higher order spatial 
derivative terms.  
In his model, 
the effects of unknown physics are modeled so as to be suppressed 
for the waves with a wavelength much 
longer than the critical scale, $k_0^{-1}$. 
Surprisingly, it was shown that 
the thermal spectrum is recovered for such modified models. 
To introduce such higher order spatial derivative terms,  
the Lorentz invariance must be violated because 
one special spatial direction needs to be chosen.  
In previous works, the rest frame of freely-falling observers 
was employed as this special reference frame. Here we give an 
extension by allowing a more general choice of the reference frame. 
Developing the method taken by Corley, 
we show that the resulting spectrum of created particles
again becomes the thermal one at the Hawking temperature 
even if the choice of the reference frame is generalized. 
Using the technique of the matched asymptotic expansion, 
we also show that the correction to the 
thermal radiation stays of order $k_0^{-2}$ or smaller when 
the spectrum of radiated particle around its peak is concerned. 
\end{abstract}


\section{Introduction}

 The thermal radiation from a black hole was first predicted by Hawking
\cite{Hawking}, which phenomenon became widely known as the 
Hawking radiation. 
This prediction is based on quantum field 
theory in curved space, which is thought of 
as an effective theory valid for low energy physics. 
However, when we consider the mechanism of the Hawking radiation, 
crucial role is played by wave packets which left the past 
null infinity with very high frequency. 
Such wave packets propagate through the collapsing 
body just before the horizon formed, and undergo 
a large redshift on the way out to the future null infinity. 
Here arises one question. Can it be justified to apply quantum 
field theory in curved space, an effective theory at 
low energy, to the phenomenon which involves 
the infinitely high frequency regime? 
There may exist some unknown physics which invalidates the 
application of the standard quantum field theory in curved space \cite{BD}. 

One of such possibilities is that the spacetime may reveal 
its discrete nature at such high frequencies. 
To take account of the effect of possible modification of 
theory in the high frequency regime, Unruh proposed a simple toy 
model by a sonic analog of a black hole \cite{Unruh,Unruh1}.  
In Unruh model, the dispersion relation of fields at high frequencies  
are modified so as to eliminate very short wavelength modes. 
In some sense, this modification is arranged to reflect the atomic 
structure of fluid which propagates sound waves. 
Usually, the group velocity of sound waves drops to 
values much less than the low frequency value when the wavelength 
becomes comparable to the atomic scale.  
In performing such modifications \cite{Unruh1}, one must assume the 
existence of a reference frame because the concept of high 
frequencies can never be a Lorentz invariant one. 
Namely, the Unruh's model manifestly breaks the Lorentz invariance. 
To our surprise, even with such a drastic change of theory, 
the spectrum observed at the future infinity was turned out to be kept 
unaltered \cite{Unruh1,JacCor,Corley}. 
Here, in this paper, we consider a generalization of this model. 

The Lorentz invariance is the very basic principle for both 
the special relativity and the general relativity. 
Hence, there are many efforts to examine the violation of the 
Lorentz invariance \cite{will}, 
and new ideas to make use of high energy 
astrophysical phenomena are also proposed recently \cite{phili}. 
However, we have not had any evidence suggesting this 
rather radical possibility yet. 
Therefore, one may think that it is not fruitful to study in detail 
such a toy model that violates the Lorentz invariance at the moment. 
But, we also have another motivation to 
study this toy model even if we could believe that the 
Lorentz invariance is an exact symmetry of the universe. 
In most of literature, the Hawking radiation was studied  
in the framework of non-interacting quantum fields in curved space.  
However, when we consider a realistic model, 
it will be necessary to consider fields with interaction terms \cite{Hoo}. 
The evolution of interacting fields 
in the background that is forming a black hole 
is a very interesting issue but to study it is very difficult. 
Hence, as a first step, it will be 
interesting to take partly into account the interaction between 
the quantum fields and the matter which is forming a black hole. 
Then, it will be natural to introduce a modified dispersion relation 
associated with the rest frame of the matter remaining 
around the black hole. 
In this sense, the Unruh's model does not require that the 
fundamental theory itself violate the Lorentz invariance. 

In order to introduce the modified dispersion relation 
we need to specify one special reference frame. 
In previous works \cite{Unruh1,JacCor,Corley,BMPS}, 
the rest frame of freely-falling observers 
was employed as the special reference frame. 
In this case, the thermal 
spectrum of the Hawking radiation was shown to be reproduced. 
However, it is still uncertain whether the same thing remains true
even when we adopt another reference frame as 
the special reference frame. 
In this paper, we give a generalization of previous 
works \cite{Unruh1,JacCor,Corley,BMPS} by allowing 
a more general choice of the reference frame. 

In most parts of the present paper, 
we follow the strategy 
taken in the paper by Corley\cite{Corley}. (See also \cite{BMPS}.) 
In his paper, as modifications of the Unruh's original model, 
two types of models were investigated. One is that with 
subluminal dispersion relation and the other is that with 
superluminal dispersion relation. It was shown
analytically that the thermal spectrum at the Hawking 
temperature is reproduced in both cases. 
However, in the superluminal case, the standard notion of 
the causal structure of black hole breaks down. 
Even if we consider 
the case that the background geometry is given by 
a Schwarzschild black hole, the wave packets 
corresponding to the radiated particles can be traced back 
to the singularity inside the horizon 
due to their superluminal nature. 
Hence, the singularity becomes naked, and 
we confront the problem of the boundary condition 
at the singularity. 
To avoid this difficulty, it is often required that the vacuum 
fluctuations be in the ground state just inside the horizon. 
However, it is not clear what is the correct boundary condition. 
As a topic related to the superluminal dispersion relation, 
it was also reported that the Hawking radiation is not necessarily 
reproduced in the models with an inner horizon \cite{CoJ}. 
In this paper, we wish to focus on the subluminal case leaving 
such a delicate issue related to the superluminal dispersion relation
for the future problem.
Even in the restriction to the subluminal case,  
it will be important to study various models 
to examine the universality of the Hawking radiation.
In the present paper, 
we finally find that the resulting spectrum of created particle
stays thermal one at the Hawking temperature 
as long as we mildly change the choice of the special reference frame. 
By a systematic use of the technique of 
the so-called matched asymptotic expansion, 
we also evaluate how small the leading correction to the 
thermal radiation is. 
On the other hand, for some 
extreme modification of the reference frame, in which case 
the analytic treatment is no longer valid, the spectrum 
is numerically shown to deviate from the thermal one significantly.   

This paper is organized as follows.
First we introduce a generalization of the Unruh's model in Sec.2.
In Sec.3 we review what quantities need to be evaluated 
in computing the spectrum of particle creation in our model. 
In Sec.4, we construct a 
solution of the field equation, and we evaluate the 
spectrum of created particles by using this solution. 
To determine the order of the leading correction to 
the thermal spectrum, we employ the method of asymptotic 
matching in Sec.5. 
In Sec.5, we also demonstrate some results of numerical calculation 
to verify our analytic results. In addition, we display  
the results for some extreme cases which are out of the 
range of validity of our analytic treatment. 
Section 6 is devoted to conclusion. 
Furthermore, 
appendix C is added to discuss the effect of scattering 
due to the modified dispersion relation. 
Although we do not think that 
this effect is directly related to the issue of Hawking radiation, 
it can in principle 
change the observed spectrum of Hawking radiation drastically if 
it accumulates throughout the long way to a distant observer.  
In the present paper, we use units with $ \hbar = c = G = 1$.

\section{Model}

In this section, we explain how we generalize the model 
that was investigated in the earlier works \cite{Unruh1,JacCor,Corley}.
Following these references, 
we consider a massless scalar field propagating in a 
2-dimensional spacetime 
given by 
\begin{equation}
ds^2=-d\tilde t^2+(d\tilde x-\tilde v(\tilde x)d\tilde t)^2, 
\label{orimet}
\end{equation}
where $\tilde v(x)$ is a function which goes to a constant 
at $\tilde x\to\infty$ and satisfies 
$\tilde v(\tilde x)\geq -1$ for 
$\tilde x\geq 0$. 
The equality holds at $\tilde x=0$. 
Since the line element $d\tilde x=0$ is null at $\tilde v=-1$,
we find that the event horizon is located at $\tilde v=-1$.
Furthermore, by the coordinate transformation given by 
$d\tilde t=d\tilde t'+\tilde v/(1-\tilde v^2)d\tilde x$, 
the above metric can be rewritten as 
\begin{equation}
ds^2=-(1-\tilde v^2)d\tilde t'{}^2+{1\over 1-\tilde v^2} 
d\tilde x^2. 
\label{met2}
\end{equation}
If we set $\tilde v(\tilde x)
=-\sqrt{2M/(\tilde x+2M)}$, this metric represents a 
2-dimensional counterpart of a Schwarzschild spacetime with 
the event horizon at $\tilde x=0$. 
In this coordinate system, the unit vector 
perpendicular to the $\tilde t=$constant hypersurfaces 
is given by $\tilde u_{\alpha}:=\partial_{\alpha} \tilde t$, and 
the differentiation in this direction is given by  
$\tilde u^{\alpha}\partial_{\alpha}=
 (\partial_{\tilde t}+\tilde v\partial_{\tilde x})$. 
We denote the unit outward pointing 
vector normal to $\tilde u_{\alpha}$ by $\tilde s^{\alpha}$. 
In order to examine 
the effect on the spectrum of 
the Hawking radiation due to a modification 
of theory in the high frequency regime, 
they investigated a system defined by the modified 
action of a scalar field, 
\begin{equation}
S={1\over 2}\int d^2\tilde x\sqrt{-g}\, g^{\alpha\beta}
{\cal D}_{\alpha}\phi^{*}{\cal D}_{\beta}\phi, 
\label{act1}
\end{equation}
where the differential operator ${\cal D}$ is defined by 
$\tilde u^{\alpha} {\cal D}_{\alpha}  =  
 \tilde u^{\alpha} \partial_{\alpha},
\tilde s^{\alpha}{\cal D}_{\alpha}  =  
 \hat F(\tilde s^{\alpha}\partial_{\alpha})$.
If we set $\hat F(z)=z$, the action (\ref{act1}) 
reduces to the standard form. 
Since we are interested in the effect caused by the change in 
the high frequency regime, we assume that 
$\hat F(z)$ differs from $z$ only for large $z$. 

In the above model, the dispersion relation 
for the scalar field manifestly breaks 
the Lorentz invariance, and there is a 
special reference frame specified by $\tilde u$. 
One can easily show that this reference frame is 
associated with a set of freely-falling observers. 
As noted in Introduction, 
it was shown that the 
spectrum of the Hawking radiation is reproduced in 
this model. 
Here we consider a further generalization of this 
model allowing to adopt other reference frames as 
the special reference frame. 

However, because of technical difficulties, we restrict 
our consideration to stationary reference frames. As we are working 
on a 2-dimensional model, the reference frame is perfectly specified
by choosing one time-like unit vector, which we denote by $u$.
Since $\partial_{\tilde t}$ in the original
coordinate system $(\tilde t, \tilde x)$ is a 
time-like Killing vector, the condition for the 
reference frame to be stationary becomes 
$\lee_{\partial_{\tilde t}} u^{\alpha}=0$,
where  $\lee_{\partial_{\tilde t}}$ is the Lee derivative in the
direction of $\partial_{\tilde t}$.  
This condition can be simply written as 
${\partial u^{\tilde \alpha}/\partial{\tilde t}}=0$, 
where we used indices associated with $\tilde{}$ to 
represent the components in the $(\tilde t,\tilde x)-$coordinate.
Furthermore, the covariant components 
$u_{\tilde \alpha}(\tilde x)
=g_{\tilde \alpha \tilde \beta}(\tilde x) 
u^{\tilde \beta}(\tilde x)$ 
is also independent of $\tilde t$.     
Thus, if we introduce a new time coordinate $t$ by
\begin{equation}
dt=u_{\tilde 0}^{-1}(\tilde x)\left( u_{\tilde 0}(\tilde x) d\tilde t+
 u_{\tilde 1}(\tilde x) d\tilde x\right)
 =d\tilde t -\gamma (\tilde x)d\tilde x,
\label{tconst}
\end{equation}
the $t-$constant hypersurfaces become manifestly perpendicular to
$u_{\alpha}$.  Here we introduced 
$\gamma(\tilde x):=-u_{\tilde 1}(\tilde x)/u_{\tilde 0}(\tilde x)$.
Further, it is convenient to choose a new spatial coordinate 
$x$ so that $\partial_t$ coincide with the Killing vector 
Since 
\begin{equation}
\partial_{t}=
 \partial_{\tilde t}+{\partial x\over \partial \tilde t} 
 \partial_{\tilde x},
\end{equation}
$x$ should be chosen as a function which depends only on $\tilde x$. 
Hence, we set 
\begin{equation}
 x :=\int_0^{\tilde x} \zeta(\tilde x') d\tilde x'. 
\label{newx}
\end{equation}
By using such a new coordinate $(t,x)$ with 
\begin{equation}
 \zeta(\tilde x')=(1-\tilde v \gamma)^2-\gamma^2, 
\end{equation}
the metric (\ref{orimet}) can be written in the form\footnote{
By considering sonic analog of the Hawking radiation, 
the use of this type of conformal metric was discussed \cite{Visser}.}, 
\begin{equation}
 ds^2= \Omega^2(x)\left[-dt^2+(dx-v(x) dt)^2\right], 
\label{newmet}
\end{equation}
where we set 
\begin{eqnarray}
 \Omega^2(x)&:=&{1\over (1-\tilde v \gamma)^2-\gamma^2}, 
\label{Omegadef}
\\
 v(x)&:=&\gamma+\tilde v(1-\tilde v \gamma). 
\label{newv}
\end{eqnarray}
Here we mention the constraint on $\Omega$. 
If we explicitly write down the condition that $u$ is a time-like 
unit vector, i.e., $u_\alpha u^\alpha=-1$, 
we find that $\Omega^2=u_{\tilde 0}^2$ holds.
Hence $\Omega^2 >0$ is guaranteed 
as long as $u$ is time-like.
Also, directly from the metric (\ref{newmet}), 
we can easily verify $\Omega^2>0$ if and only if 
the $t-$constant hypersurfaces are space-like. 
Hence, it will be appropriate to assume $\Omega^2>0$. 
Then, we find that $\Omega^2$ has a finite minimum value 
when $|\tilde v|<1$. 
The minimum value is $1-\tilde v^2$ 
which is realized when $v=0$. 

Next we write down the explicit form of $u$ and $s$. 
By using the facts that $u$ is perpendicular to 
the $t-$constant hypersurfaces and that it is an unit vector, 
we can show that 
the differentiation in the direction of $u$ 
is given by 
\begin{equation}
u^{\alpha}\partial_{\alpha}
={1\over{\Omega}}\,(\partial_t+v\partial_x). 
\label{equ}
\end{equation}
Similarly, we can show that
the differentiation in the direction of $s$, which is an unit vector 
perpendicular to $u$, is given by 
\begin{equation}
s^{\alpha}\partial_{\alpha}={1\over{\Omega}}\,{\partial_x}.
\end{equation}
By using $u$ and $s$, the metric can be represented as 
$g^{\alpha\beta}=-u^{\alpha}u^{\beta}+s^{\alpha}s^{\beta}$,  and 
the determinant of $g_{\alpha\beta}$ becomes $g=- \Omega ^4$.

Now, we find that to generalize the choice of the 
special reference frame introduced to set the 
modified dispersion 
relation is equivalent to generalize 
the metric form given in (\ref{orimet}) to 
the one given in (\ref{newmet}) replacing 
$\tilde u$ and $\tilde s$ with $u$ and $s$ 
in the defining equations of the differential operator ${\cal D}$. 
As a result, the action (\ref{act1}) becomes
\begin{equation}
S={1\over 2}\int dtdx \left[ |(\partial_t+v\partial_x)\phi|^2-
{{\Omega}(x)}^2 \left|{\hat F}
\left({1\over{\Omega}}\partial_x \right) \phi \right|^{2} \right].
\label{action2}
\end{equation}
If we set $\Omega^2\equiv 1$, 
the models are reduced to the original one discussed 
in Ref.~\cite{Corley}. 

Here, we should mention one important relation for the later 
use. The temperature of the Hawking spectrum is determined 
by the surface gravity $\kappa$ defined 
by $\kappa:=d\tilde v/d\tilde x|_{\tilde x=0}$. 
The surface gravity $\kappa$ can also be represented as \cite{JacKan} 
\begin{equation}
\kappa=\left.{dv\over dx}\right\vert_{x=0}.
\label{temperature}
\end{equation}
In order to verify this relation, 
we evaluate $dv/dx$ by using Eqs.~(\ref{newx}) and (\ref{newv}) to obtain 
\begin{equation}
{dv\over dx}={1\over {(1-\tilde v \gamma)^{2}-\gamma^{2}}}
(\gamma_{,\tilde x}+\tilde v_{,\tilde x}
(1-\tilde v\gamma)-\tilde v\tilde v_{,\tilde x} \gamma
  -\tilde v^{2}\gamma_{,\tilde x}).
\label{surface}
\end{equation}
From Eq.~(\ref{newv}), we also find $v=-1$ when $x=0$. 
Hence, substituting $v=-1$ into 
(\ref{surface}), we obtain Eq.~(\ref{temperature}).

Then, let us derive the field equation by taking the 
variation of the action (\ref{action2}). 
Assuming that ${\hat F}(z)$ is an odd function of $z$,  
we obtain 
\begin{equation}
({\partial}_{t}+{\partial}_{x}v)({\partial}_{t}+v{\partial}_{x}){\phi}
={\Omega}{\hat F}\!\left({1\over{\Omega}}{\partial}_{x} \right)
 {\Omega}
\,{\hat F}\!\left({1\over{\Omega}}{\partial}_{x} \right){\phi}.
\label{field equation}
\end{equation}

To proceed further calculations, 
we need to assume a specific dispersion relation. 
Following Ref.~\cite{Corley}, we adopt here 
\begin{equation}
\hat F(z)=z+{1\over 2k_0^2} z^3, 
\label{dispersion}
\end{equation}
where $k_0$ is a constant. 
Since the model should be arranged to differ from the 
ordinary one only in the high frequency regime, 
the critical wave number $k_0$ is supposed to 
be sufficiently large. 
With this choice of $\hat F$, 
neglecting the terms that are inversely proportional to the 
fourth power of $k_0$, the field equation becomes
\begin{equation}
({\partial}_{t}+{\partial}_{x}v)({\partial}_{t}+v{\partial}_{x}){\phi}
=\left[\partial_x^2+{1\over 2 k_0^2}
\left( \partial_x^2 {1\over \Omega} \partial_x {1\over \Omega}
 \partial_x \right)+
{1\over 2 k_0^2} \left(\partial_x{1\over \Omega} \partial_x
{1\over \Omega} \partial_x^2\right)\right] \phi. 
\label{feq}
\end{equation}

Before closing this section, we briefly discuss 
the meaning of the functions $\Omega$ and $v$. 
From (\ref{equ}), it is easy to understand that 
$v$ is the coordinate velocity of the integration curves of $u$. 
To understand the meaning of $\Omega$, we further calculate 
the covariant acceleration of the 
integral curves of $u$, 
$|u^{\alpha}_{~;\beta} u^{\beta}|$, where semicolon represents 
the covariant differentiation. 
After a straightforward calculation, 
we see that the covariant acceleration is given by 
$|\partial_x \Omega^{-1}|$. 
Hence, we find that the derivative of $\Omega^{-1}$ gives 
the acceleration of the reference frame.

\section{particle creation rate}

In this section, we briefly review how to evaluate the 
spectrum of Hawking radiation. We clarify what quantities 
need to be calculated for this purpose. 
(For the details, see Ref.~\cite{JacCor}.)

To evaluate the spectrum of Hawking radiation, we need 
to solve the field equation (\ref{feq}) with an appropriate 
boundary condition. 
However, owing to the time translation invariance with the 
Killing vector $\partial_t$, 
we do not have to solve the partial differential 
equation (\ref{feq}) directly. 
By setting 
$ {\phi}(t,x)=e^{-i{\omega}t}{\psi}(x)$, 
Eq.~(\ref{feq}) reduces to an ordinary differential 
equation (ODE), 
\begin{equation}
\left[(-i\omega+{\partial}_{x} v)(-i\omega +v{\partial}_{x})
-\left\{{{\partial}_x}^{2}+{1\over {2 {k_0}^2}}
{\left({{\partial}_{x}}^{2}{1\over{\Omega}}
 {\partial}_{x}{1\over{\Omega}}
{\partial}_{x}\right)}+
{1\over{2 {k_0}^2}}{\left({\partial}_{x}{1\over{\Omega}}{\partial}_{x}
{1\over{\Omega}}{{\partial}_{x}}^2\right)}\right\}\right]
\psi =0. 
\label{ode} 
\end{equation} 
We could not make use of this simplification if we 
relax the restriction that the reference frame 
should be stationary. 
 
Here we note that the norm of 
$\partial_t$ is given by $|\partial_t|=\sqrt{1-v^2}\Omega=
\sqrt{1-\tilde v^2}$.
Therefore, the frequency $\omega^{(stat)}$ for the 
static observers who stay at a constant
$x$(or $\tilde x$) is related to $\omega$ by 
\begin{equation}
\omega^{(stat)}=\omega/\sqrt{1-\tilde v^2}.
\label{omegas}
\end{equation}
Hence, as long as $\tilde v_{\infty}:=\tilde v(x\to\infty)$
is not equal to zero, $\omega^{(stat)}$ differs from $\omega$. 
By looking at the metric (\ref{met2}) in the static chart, 
we find that this frequency shift is merely 
caused by the familiar effect due to the gravitational redshift.
Therefore, even if we consider models with $\tilde v_{\infty}\ne 0$, 
$\omega$ might be identified with the frequency observed at 
the hypothetical infinity where the gravitational potential 
is set to be zero. 
However, the situation is more transparent 
if we can set $\tilde v_{\infty}=0$ 
like the 2-dimensional black hole case. 
In this case, we can identify $\omega$ with $\omega^{(stat)}$ 
without any ambiguity. 
As mentioned in Ref. \cite{JacCor}, there is a difficulty in evaluating 
the spectrum of radiation in the case of $v_{\infty}:=v(x\to\infty)=0$.  
In previous models, $v_{\infty}=0$ directly implies $\tilde v_{\infty}=0$. 
Therefore, we could not apply the result to  
the example of a 2-dimensional black hole spacetime directly\footnote{
In the model proposed in Ref.~\cite{lattice}, the case with $v_{\infty}=0$
can be dealt with.}. 
On this point, in our extended model, 
the cases with $\tilde v_{\infty}=0$ can be dealt with 
since $\tilde v_{\infty}=0$ does not mean 
$v_{\infty}\ne 0$. 

Let us return to the problem to solve Eq.~(\ref{ode}).
From the above ODE, 
the asymptotic solution at $x\to\infty$ is easily obtained 
by assuming the plane wave solution like
\begin{equation}
\psi(x)\propto e^{ikx}. 
\end{equation}
Substituting this form into Eq.~(\ref{ode}), 
we obtain the dispersion relation 
\begin{equation}
(\omega - v_{\infty}k)^{2} = k^{2} - {k^{4} \over k_0^{2} \Omega_{\infty}^2},
\label{disp}
\end{equation}
where 
$\Omega_{\infty}$ is the asymptotic 
constant value of $\Omega$. 
The quantity on the left hand side 
\begin{equation}
\omega':=\omega-v k,
\end{equation}
is related to the frequency measured by the observers 
in the special reference frame. In fact, 
this frequency divided by $\Omega$ is the factor that appears 
when we act the operator $iu^{\alpha}\partial_\alpha$ 
on a wave function $e^{-i(\omega t-k x)}$.
As shown in Ref.~\cite{JacCor}, two of the 4 solutions of Eq.~(\ref{disp}) 
have large absolute values, which we denote by $k_{\pm}$, 
and the other two have small absolute values, which we 
denote $k_{\pm s}$. For each pair, one is positive and 
the other is negative. 
The subscript $\pm$ represents 
the signature of the solution. 
Then, the general solution of Eq.~(\ref{disp}) at $x\to\infty$ is 
given by a superposition of these plane wave solutions as
\begin{equation}
\psi(x)=\sum_{l=\pm,\pm s} c_{l}(\omega)e^{ik_l(\omega)x}.
\end{equation}

By such a local analysis, however, 
the coefficients $c_{\ell}(\omega)$ 
are not determined. 
To determine them, we need to find 
the solution that 
satisfies the boundary condition 
corresponding to no ingoing waves plunging into the event horizon. 
This condition is slightly different from the condition that 
the solution of ODE~(\ref{ode}) vanishes inside the horizon. 
The latter condition is stronger than the former one 
because the latter one also prohibits the pure outgoing wave
from the event horizon which may exist in the present model  
with the modified dispersion relation. The former 
condition is the sufficient condition to determine the 
wave function uniquely, while the existence of the 
solution that satisfies the latter condition 
is not guaranteed in general. 
However, once we find such a solution that satisfies 
the latter stronger condition, it is the solution 
that satisfies the required boundary condition. 
In the succeeding sections, we solve ODE~(\ref{ode}) 
requiring the latter condition. 

{}Finally, we present the formula 
to evaluate the expectation value of 
the number of emitted particles 
naturally defined at $x\to\infty$. 
In spite of our generalization of models, 
the same derivation of the formula that is given in Ref.~\cite{JacCor} 
is still valid. 
The same arguments follow without any change, but one 
possible subtlety exists on the point 
whether the expression 
of the conserved inner product is unaltered or not. 
Therefore, we briefly explain this point. 
The defining expression for the conserved inner product 
given in Ref.~\cite{JacCor} is 
\begin{equation}
({\phi_1},{\phi_2})=i{\int}dx[{\phi_1}^{*}
         ({\partial}_{t}+v{\partial}_{x})
{\phi_2}-{\phi_2}({\partial}_{t}+v{\partial}_{x}){\phi_1}^{*}], 
\label{inner}
\end{equation}
where the integration is taken over a $t$-constant hypersurface.
Here both ${\phi_1}$ and ${\phi}_2$ are supposed to be 
solutions of the field equation. 
The constancy of this inner product is 
related to the invariance of the function 
\begin{equation}
{\cal L}(\phi_1,\phi_2):=(\partial_t +\partial_x)\phi_1 \cdot
(\partial_t +\partial_x)\bar\phi_2
-\Omega^2 \left[\hat F\left({1\over \Omega}\partial_x \right)\phi_1\right]
\cdot \left[\hat F\left({1\over \Omega}\partial_x \right)\bar\phi_2\right], 
\end{equation}
under the global phase transformation
\begin{equation}
\phi\to e^{i\lambda}\phi.
\end{equation} 
Taking the differentiation of 
${\cal L}(e^{i\lambda}\phi_1,e^{i\lambda}\phi_2)$ 
with respect to $\lambda$, we have
\begin{eqnarray}
 0={d{\cal L}(e^{i\lambda}\phi_1,e^{i\lambda}\phi_2)
\over d\lambda}&=&{\partial {\cal L}\over \partial\phi_1}i\phi+
       {\partial {\cal L}\over \partial\dot\phi_1}{d(i\phi_1)\over dt}
       +{\partial {\cal L}\over \partial\bar\phi_2}(-i\bar\phi_2)+
  {\partial {\cal L}\over \partial\dot{\bar\phi_2}}{d(-i\bar\phi_2)\over dt}
\cr
 &=&i{d\over dt}\left({\partial {\cal L}\over \partial\dot\phi_1}\phi_1-
   {\partial {\cal L}\over \partial\dot{\bar \phi_2}}
        {\bar\phi_2}
 \right), 
\label{constip}
\end{eqnarray}
where we used the field equation in the last equality. 
Eq.~(\ref{constip}) proves the constancy of the inner 
product (\ref{inner}). 
Of course, the constancy of the inner product (\ref{inner}) 
can also be verified by directly calculating its 
$t$-derivative using the field equation. 

Now that we verified that the extension of model 
does not alter the expression of the inner product, 
we just quote the formula from Ref.~\cite{JacCor}.   
For a wave packet which is peaked around a frequency $\omega$, 
the expectation value of the number of created 
particles is given by 
\begin{equation}
N(\omega)={|\omega^{'}(k_-) v_g(k_-) c_-^{2}(\omega)|
\over{|\omega^{'}(k_{+s}) v_g(k_{+s}) c_{+s}^{2}(\omega)|} }. 
\label{number}
\end{equation}
where $ v_g(k):=\partial\omega(k)/\partial k$ 
is the group velocity measured by a static 
observer. 

\section{solving field equation}
In this section, 
to determine the coefficients $c_l$,
we solve the field equation (\ref{ode}) by using 
several approximations. 
In the region close to the horizon, we use 
the method of Fourier transform.  
The solution is found to be uniquely determined by imposing the 
boundary condition discussed in the preceding section. 
On the other hand, in the region sufficiently far 
from the horizon, we construct four independent solutions 
which become $e^{ik_{\ell}(\omega) x}$ at $x\to \infty$. 
We use the WKB approximation for the two short-wavelength 
modes and we use the simple $1/k_0^2$ expansion 
for the other two long-wavelength modes. 
Later, we find that these two different regions of validity 
have an overlapping interval 
as long as $k_0$ is taken to be sufficiently large.  
Hence, the requirement that the solutions obtained in both regions 
match in this overlapping interval 
determines the coefficients $c_{\ell}(\omega)$. 

Basically, our computation is an extension 
of that given by Corley~\cite{Corley}. 
Here we take into 
account the generalization of models discussed in Sec.~2. 
Furthermore, to evaluate the order of the leading deviation 
from the thermal spectrum, 
we shall take into account some higher order terms. 
At the same time, we also carefully keep counting 
the order of errors contained in our estimation. 
For brevity, 
we concentrate on the most interesting 
case in which $\omega$ and $\kappa$ are same order. 

\subsection{the case close to the horizon}

Now we want to find a solution satisfying the boundary condition 
that the wave function rapidly decrease in the horizon.   
Therefore, we restrict our consideration to the region 
close to the horizon, $|x|<x_1$, by choosing a sufficiently small $x_1$.   
We introduce a parameter, 
\begin{equation}
 \delta:=\max_{|x|<x_1, i=0, 1, 2} \tilde\delta_i(x),  
\label{ddelta}
\end{equation}
with 
\begin{equation}
 \tilde\delta_0:=\kappa x, \quad 
 \tilde\delta_1:={\kappa_1^2 x\over \kappa},  \quad 
 \tilde\delta_2:={\Omega_0\over \Omega_1} x.
\end{equation}
Since we wish to think of $\delta$ as a small parameter 
for the perturbative expansion, we 
require $\delta\ll 1$. 
Then, we find that $x_1$ must be chosen to satisfy 
\begin{equation}
 x_1\ll \mbox{min}\left(1/\kappa, \kappa/\kappa_1^2, 
   \Omega_1/\Omega_0\right).
\label{x_1}
\end{equation}
We assume that $v(x)$ and $1/\Omega(x)$ can be 
expanded around the horizon like 
\begin{eqnarray}
v(x) & = & -1+\kappa x+{1\over 2} \kappa_1^2 x^2+O(\delta^3), 
\cr
\left({1\over \Omega}\right) & \approx & {1\over \Omega_0} 
 +{1\over \Omega_1} x +{1\over \Omega_0}O(\delta^2). 
\label{vomegat}
\end{eqnarray}
Substituting these expansions into the field equation (\ref{ode}), 
we classify the terms into five parts 
according to the number of differentiations 
acting on $\psi$.
Then, keeping the leading order correction terms 
with respect to $\delta$ in each part, 
we obtain the field equation valid in the region 
close to the horizon as 
\begin{eqnarray}
L[\psi(x)]: = 
&& {1\over k_0^2} \left[{1\over \Omega_0^2}
+{2\over \Omega_0 \Omega_1}x\right]\psi^{(4)}
+{4\over k_0^2}{1\over \Omega_0 \Omega_1} \psi'''
+(2\kappa x -\kappa^{2} x^{2}+\kappa_1^2 x^2) \psi''
\nonumber\\ 
&& + 2\left[ -(i\omega - \kappa)+(\kappa (i \omega - \kappa ) +
  \kappa_1^2)x \right] \psi'
- i \omega (i \omega - \kappa - \kappa_1^2 x) \psi = 0.  
\label{horizon feq}
\end{eqnarray}
In Corley's paper \cite{Corley}, the terms of 
$O(\delta^{1})$ were neglected, 
while we keep them in the present paper.

Here, we introduce 
the momentum-space representation of the 
wave function, $\psi(s)$, by 
\begin{equation}
\psi(x)=\int_{C}{\, }ds{\, }e^{sx} \hat{\psi}(s).  
\label{int}
\end{equation} 
Substituting this expression into Eq.~(\ref{horizon feq}), 
we perform the integration by parts like 
\begin{eqnarray*}
\int_C ds\,  x e^{sx}\hat\psi(s)=\int_C ds\, 
 \left({\partial\over\partial s} e^{sx}\right)
 \hat\psi(s)= - \int_C ds\, e^{sx}
 \left({\partial\over\partial s} \hat\psi(s)\right)
 +[e^{sx} \hat\psi(s)]_{s_i}^{s_f},
\end{eqnarray*}
where $s_i$ and $s_f$ are the start and the end points 
of integration, respectively. 
Note that there appear surface terms like 
the last term on the right hand side in the above example. 
Then, we find that the field equation in the momentum space is
given by 
\begin{equation}
L[\psi(x)]:=\int_{C}{\, }ds{\, } {\hat L}[\hat \psi(s)]
+ \mbox{(boundary terms)} = 0, 
\label{fourier}
\end{equation}
where
\begin{eqnarray}
{\hat L}[\hat \psi(s)]:=&&
{1\over k_0^2\Omega_0^2}\left(1-
  2{\Omega_0\over \Omega_1}\left(\partial_s-{2\over s}\right)\right) 
      s^4\hat\psi 
-\left(2\kappa \partial_s +(\kappa^2+\kappa_1^2)\partial_s^2
   \right) s^2\hat\psi
\nonumber\\ &&
+2\left[ -(i\omega - \kappa)-(\kappa(i\omega-\kappa)+
\kappa_1^2)\partial_s \right]s \hat\psi
-i\omega(i\omega-\kappa+\kappa_1^2 \partial_s)\hat\psi. 
\label{bt}
\end{eqnarray}

If we are allowed to neglect the boundary terms, 
we can construct a solution of Eq.~(\ref{horizon feq}) 
by using Eq.~(\ref{int}) 
from a solution $\hat \psi(s)$ which satisfies
\begin{equation}
{\hat L}[\hat \psi(s)]=0, 
\label{bt1}
\end{equation}
In the following, we solve Eq.~(\ref{bt1}) without taking 
any care about whether the boundary terms can be neglected or not. 
After we find a solution, we verify that the corresponding boundary terms 
are sufficiently small.

To solve Eq.~(\ref{bt1}), 
it is convenient to introduce a new variable
\begin{equation}
 W:={\partial \ln(s^2\hat\psi)\over \partial \ln s}.
\label{W1}
\end{equation}
Taking $\delta$ as a small parameter, we expand $W$ as 
\begin{equation}
 W=W^{(0)}+W^{(1)}+O(\delta^2).   
\label{Wexpansion}
\end{equation}
Then, $W^{(0)}$ is found to be given by 
\begin{equation}
W^{(0)}
 = \left[{\tilde\epsilon^2\over 2}(sx)^3
    +\left(1-i {\omega\over\kappa}\right)\right],  
\label{s1}
\end{equation}
and $W^{(1)}$ is given by
\begin{equation}
 W^{(1)} = {\cal R}(W^{(0)}), 
\end{equation}
with 
\begin{eqnarray}
 {\cal R}(W) & := & 
 \tilde\delta_0 \left[ -{1\over 2}{\partial W\over \partial \ln s}
  -{1\over 2} W^2+{1\over 2}W +
  \left(1-{i\omega\over \kappa}\right)
  \left(W+{i\omega\over 2\kappa}-1\right) \right](sx)^{-1}
 -\tilde\delta_2 \tilde\epsilon^{2} W (sx)^2
\cr && 
 -\tilde\delta_1
    \left( -{1\over 2}{\partial W\over \partial \ln s}-{1\over 2} W^2
   +{3\over 2}W -1 \right)(sx)^{-1}.  
\end{eqnarray}
Here we defined
\begin{equation}
\tilde\epsilon:=\sqrt{1\over k_0^2\Omega_0^2\kappa x^3}.  
\label{ep2}
\end{equation}
Although we later consider the situation in which 
$\tilde\epsilon$ is small, 
$\tilde\epsilon$ is not small at all in the region close to the horizon. 
Substituting the explicit expression for $W^{(0)}$ 
into ${\cal R}(W)$, we obtain 
\begin{eqnarray}
W^{(1)} &=& 
  {\tilde\epsilon^4 (-\tilde\delta_0+\tilde
\delta_1-4\tilde\delta_2) \over 8}(sx)^{5}
  -{\tilde\epsilon^2 \over 2}\left(\tilde\delta_0 
+(\tilde\delta_1-2\tilde\delta_2) 
   \left({i \omega \over \kappa}-1 \right) \right)(sx)^{2} 
\cr &&
 + {i \omega \over {2 \kappa}}
\tilde\delta_1 \left( {i \omega \over \kappa} + 1 \right)(sx)^{-1}. 
\end{eqnarray}
Under the condition that the boundary terms in Eq.~(\ref{fourier}) 
can be neglected, 
we obtain the solution 
$\psi(x)$ which is valid up to $O(\delta^1)$ as 
\begin{equation}
\psi(x) \approx \int_{C} ds\,  e^{xf(s)}, 
\label{int2}
\end{equation}
where,
\begin{equation}
xf(s) = x  f_0(s) + x f_1(s),  
\label{xfs}
\end{equation}
is given by
\begin{eqnarray}
x f_0(s) &:=& x s + \int {ds\over s} W^{(0)} -2\ln s
\cr &=&  xs +{ \tilde\epsilon^2 (xs)^{3} \over 6}
+\left(-1-{iw\over \kappa}\right)\ln s, 
\label{xf0}
\\
x  f_1(s)&:=& \int {ds\over s} W^{(1)}
\cr &=& {\tilde\epsilon^4 (-\tilde\delta_0
+\tilde\delta_1-4\tilde\delta_2) \over 40} (sx)^{5}
 -{\tilde\epsilon^2 \over 4}
\left(\tilde\delta_0+(\tilde\delta_1-2 \tilde\delta_2) 
\left({i \omega \over \kappa}-1 \right)\right)(sx)^{2} 
\cr &&
  - {i \omega \over {2 \kappa}}
\tilde\delta_1 \left( {i \omega \over \kappa} + 1 \right)(sx)^{-1}. 
\label{xf1}
\end{eqnarray} 

The boundary condition of the wave function 
requires that it exponentially decreases 
inside the horizon.  
In order to construct the wave function that satisfies 
this boundary condition, we must choose the contour 
of integration appropriately. 
We propose to adopt the contour $C$ given in Fig.~1. 
This contour does not go to infinity but have end points 
(denoted by open circles in Fig.~1) 
at which the absolute value of $s$ is sufficiently large 
but does not exceed the limit given in (\ref{region1}). 
Hence, as shown in Appendix \ref{appendixA}, 
the higher order correction does not dominate along 
this contour. 
Furthermore, 
as is also explained in Appendix \ref{appendixA}, 
with this choice of end points, 
the boundary terms in Eq.~(\ref{fourier}) 
are exponentially small, and can be neglected. 

\begin{figure}[tbh]
\vspace*{0cm}
\centerline{\epsfxsize5cm \epsfbox{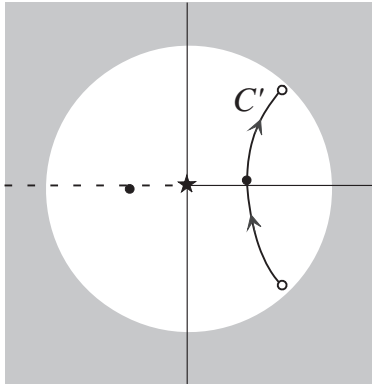}}
\caption{The contour for integration is chosen for the 
correction terms not to dominate the leading terms. 
In the shaded region, the higher order terms with respect to $\delta$ 
becomes larger than the leading terms. Hence, the contour is 
shortened so as not to violate the validity of approximation.  
}
\label{fig1}
\end{figure}

Next we show that $\psi(x)$ given by (\ref{int2}) is 
actually the solution that satisfies the required boundary condition.
In order to evaluate the integration (\ref{int2}) analytically,  
we use the method of steepest descents.  
For this method to be valid,  
$1/|x s_0|\ll 1$ is required, where $s_0$ is the 
value of $s$ at the saddle point that dominates the integral.  
Then, this requirement can be rewritten as 
\begin{equation}
|x|> x_2, 
\end{equation}
where we must choose $x_2$ to satisfy $(1/|x s_0|)_{x=x_2}\ll 1$. 
We shall see later that $1/|x s_0|$ is of $O(|\tilde\epsilon|)$. 
Hence, for the first time at this moment, we further restrict 
our consideration to the region in 
which $\tilde\epsilon$ is also small. 
In order that the condition $|x|>x_2$ to be 
compatible with $|x|<x_1$, 
\begin{equation}
\left({\kappa\over k_0\Omega_0}\right)^{2/3}
\ll \mbox{min}(1, \kappa^2/\kappa_1^2, \kappa\Omega_1/\Omega_0), 
\label{matchcondi}
\end{equation}
is required. 
However, this requirement to the model parameters 
will not reduce the generality of 
our analysis so much 
because we are interested in the case 
that the typical length-scale for the modification of dispersion 
relation, $k_0^{-1}$, is sufficiently small.

As in the case of $\delta$, we introduce an expansion parameter 
\begin{equation}
 \epsilon:=|\tilde\epsilon(x_2)|, 
\end{equation}
and we neglect the terms that induce the relative error of  
$O(\epsilon^2)$ or smaller in the amplitude of the wave function. 
As for $\delta$, we also keep the terms up to linear order 
in $\delta$. 
Here one remark is in order.  We 
imposed a further restriction $|x|>x_2$ 
to evaluate the explicit form of the solution (\ref{int2}). 
We stress, however, that 
the solution (\ref{int2}) itself is valid throughout the 
region $|x|<x_1$.

To evaluate (\ref{int2}) by using 
the method of steepest descents,  we 
need to know the value of $s$ at saddle points which are 
determined by solving 
\begin{eqnarray}
 f'(s) & \equiv & f'_0(s)+f'_1(s) 
\cr 
&=& \left[1+{\tilde\epsilon^{2}(sx)^{2}\over{2}}
-\left(1+{i\omega \over{\kappa}}\right)(sx)^{-1}\right]
 +(sx)^{-1}W^{(1)}=0. 
\end{eqnarray}
We solve this equation by assuming that the solution is 
given by a power series expansion with respect to $\delta$ as
\begin{equation}
s_{\pm } = s_{0\pm}
  + s_{1\pm}+ s_{2\pm}\cdots.   
\end{equation}
For our present purpose,
it is enough to find the solution in the form 
of series expansion with respect to $\epsilon$.  
One solution of $x s_0$ is of $O(1)$, and 
the integration along the path through this saddle point 
cannot be evaluated by the method of steepest descents. 
The other two solutions are given by 
\begin{equation}
 x s_{0\pm}:=\mp \left(\sqrt{-\tilde\epsilon^{2}\over 2}\right)^{-1} 
   -{1+i\omega/\kappa \over 2}
   \pm \sqrt{{-\tilde\epsilon^{2}\over 2}}
   {3(1+i\omega/\kappa)^2\over8}
   +O(\epsilon^{2}),
\label{sai}
\end{equation}
and 
\begin{eqnarray}
xs_1 = &\pm& \left(\sqrt{-\epsilon^2\over 2}\right)^{-1} 
\left({\delta_0\over 4}-{\delta_1\over 4}+ \delta_2\right)
+{1\over 4}\left[\left(3+{i\omega\over \kappa}\right)\delta_{0}+
 \left(-3+{i\omega\over\kappa}\right)\delta_{1}
 +8\delta_{2}\right]
\cr & \mp & 
{1\over 32}\sqrt{-\epsilon^2\over 2}\left(1+{i\omega\over \kappa}\right)
\left[\left(9+{i\omega\over \kappa}\right)\delta_{0}
-3\left(3-{5i\omega \over \kappa}\right)\delta_{1}
+4\left(5-{3i\omega \over \kappa}\right)\delta_{2}
\right]+O(\epsilon^{2}). 
\label{anten2p}
\end{eqnarray}
Since $ |s_{max} / s_{0\pm}| \sim \sqrt{1/\tilde  \delta} \gg 1$,
these saddle points are contained in the region in which the expansion
with respect to $\delta$ is valid.
In the following,  to keep the notational simplicity,  
we abbreviate the subscript 
$\pm$ from $s_{0\pm}$ and $s_{1\pm}$ unless it causes any ambiguity.  

Now we evaluate the integration (\ref{int2}) by
using the method of steepest descents. 
For the contour given in Fig.~1,  only the saddle point $s_{+}$
dominantly contributes to the integration inside the horizon. 
For our present purpose, the formula  
\begin{equation}
\psi(x) \approx 
{-\sqrt{2\pi}e^{xf(s_{\pm})}\over{(-xf''(s_{\pm}))^{1/2}}}
\left[1-{5\over{24}}{(xf'''(s_{\pm}))^{2}\over{(xf''(s_{\pm}))^{3}}}
+{1\over 8}{xf^{(4)}(s_{\pm})\over {(xf''(s_{\pm}))^{2}}}
+O(\epsilon^2)\right],
\label{anten3}
\end{equation}
is accurate enough to keep  
the correction up to $O(\epsilon^{1})$. 
The details of calculation to evaluate Eq.~(\ref{anten3}) 
up to $O(\epsilon, \delta)$ is given in 
Appendix \ref{appendixB}. 
In the end, we obtain 
\begin{eqnarray}
\psi(x) &\approx &
\sqrt{2 \pi \kappa}
(k_0 \Omega_0)^{-{i \omega \over \kappa}-{1\over 2}}
(-2 \kappa x)^{- {3 \over 4} - {i \omega \over {2 \kappa}}}
\exp \left(-{2 \over 3} \sqrt{2\kappa}k_0 
\Omega_0 (-x)^{3/2}
+W + O(\epsilon^{2} ,  \delta^{2})
\right),
\label{fsolution}
\end{eqnarray}
where 
\begin{eqnarray}
W&=&
-\left(\sqrt{-\tilde \epsilon^2\over 2}\right)^{-1}\left(
-{1\over 10}\tilde\delta_0
+{1\over 10}\tilde\delta_1
-{2\over 5}\tilde\delta_2\right)
\cr &&
+\left({i\omega \over {4 \kappa}}+{3\over8}\right)\tilde\delta_0
+\left({i\omega \over 4\kappa} - {3\over 8}\right)\tilde \delta_1
+{1\over 2}\tilde \delta_2 
\cr &&
+\sqrt{{-\tilde\epsilon^{2} \over 2}}
\Biggl[\left(-{41\over 48}-{i\omega\over \kappa}
-{1\over 4}\left({i\omega\over \kappa}\right)^2\right)
+\left({11\over 64}+{1\over 4}{i\omega\over \kappa}
+{1\over 16}\left({i\omega\over \kappa}\right)^2\right)\tilde\delta_0
\cr && \quad\quad
+\left(-{11\over 64}+{3\over 4}{i\omega\over \kappa}
+{15\over 16}\left({i\omega\over \kappa}\right)^2\right)\tilde\delta_1
+\left(-{5\over 16}-{i\omega\over \kappa}
-{3\over 4}\left({i\omega\over \kappa}\right)^2\right)\tilde\delta_2
\Biggr].
\label{name2}
\end{eqnarray}
We can immediately see that amplitude of the wave function 
reduces exponentially as we decrease $x$ (as we increase $-x$).

\begin{figure}[tbh]
\vspace*{0cm}
\centerline{\epsfxsize5cm \epsfbox{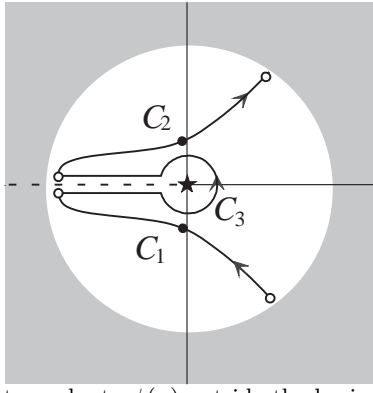}}
\caption{The deformed integration contour to evaluate $\psi(x)$ 
outside the horizon. In $x>0$ region, the saddle points 
move to the neighborhood of the imaginary axis. The 
contours $C_1$ and $C_2$ are chosen to pass through these 
saddle points and to be able to evaluate the integrations 
along them by the steepest descents. The remaining part 
of the integration contour which goes 
around the branch cut is called $C_3$. 
}
\label{fig3}
\end{figure}

Next we turn to evaluate $ \psi (x) $ for $ x > 0 $. 
We use the method of steepest descents again. 
In the present case, 
the location of saddle points moves to points on the
imaginary axis on the complex $s-$plane. 
The leading order approximation is given by 
\begin{equation}
xs_{0 \pm} = \pm {\sqrt{2} i\over \epsilon_2} + \cdots.  
\label{sao}
\end{equation}
Therefore, to evaluate (\ref{int2}) by using the method of 
steepest descents, we need to deform the contour of integration. 
At this point, we must take account of the 
existence of a branch cut emanating from $s=0$, 
which originates from the log-term in the integrand.
We choose this branch cut along the negative side of the real axis.
Then, deforming the contour so as to go through these two saddle points 
we find that the contour is divided into three pieces 
as shown in Fig.~\ref{fig3}.  
We respectively denote them by $C_1 , C_2$ and $C_3$. 
$C_1$ and $C_2$ are the contours passing through the 
saddle points $s_-$ and $s_+$, respectively. 
Both contours have a new boundary point which 
is chosen to satisfy $|s|<s_{max}$. 
The contour $C_3$ connects these two newly 
introduced boundary points, going around the origin 
in the anti-clockwise manner. 

{}First we evaluate the integrations along the contours $C_1$ and $C_2$ 
by using the method of steepest descents. 
Just repeating the same calculation as in the case of $x<0$, 
these integrations are evaluated as 
\begin{equation}
\psi_{1,2} (x) = 
e^{\mp \pi \omega / 2 \kappa }{ 1 \over { \sqrt{ k_0 \Omega_0} }} 
(k_0 \Omega_0)^{-{i \omega \over \kappa}}
\sqrt{2 \pi \kappa}
(2 \kappa x)^{ - {3 \over 4} - {i \omega \over {2 \kappa}}}
\exp \left( \mp i {2 \over 3} \sqrt{2 \kappa}
k_0 \Omega_0 x^{3/ 2}
+W_{1,2}
+O(\epsilon^{2},\delta^{2})\right),
\label{near12}
\end{equation}
where 
\begin{eqnarray}
W_{1,2}&=&
\mp i {\sqrt{2} \over \tilde \epsilon}\left(
-{1\over 10}\tilde\delta_0
+{1\over 10}\tilde\delta_1
-{2\over 5}\tilde\delta_2\right)
\cr &&
+\left({i\omega \over {4 \kappa}}+{3\over8}\right)\tilde\delta_0
+\left({i\omega \over 4\kappa} - {3\over 8}\right)\tilde \delta_1
+{1\over 2}\tilde \delta_2 
\cr &&
\mp i {\tilde\epsilon\over \sqrt{2}}
\Biggl[\left(-{41\over 48}-{i\omega\over \kappa}
-{1\over 4}\left({i\omega\over \kappa}\right)^2\right)
+\left({11\over 64}+{1\over 4}{i\omega\over \kappa}
+{1\over 16}\left({i\omega\over \kappa}\right)^2\right)\tilde\delta_0
\cr && \quad\quad
+\left(-{11\over 64}+{3\over 4}{i\omega\over \kappa}
+{15\over 16}\left({i\omega\over \kappa}\right)^2\right)\tilde\delta_1
+\left(-{5\over 16}-{i\omega\over \kappa}
-{3\over 4}\left({i\omega\over \kappa}\right)^2\right)\tilde\delta_2
\Biggr].
\label{name3}
\end{eqnarray}

Next we consider the integral along the $C_3$ contour.
Here, we divide $xf(s)$ given in Eq.~(\ref{xfs}) 
into two parts as 
\begin{eqnarray}
&&
x \bar f_0(s):= xs + \left(-1-{i \omega \over \kappa} \right) \ln s
\cr &&
x \bar f_1(s):= {\tilde\epsilon^2 (xs)^3\over 6}+xf_1(s),
\label{newsep}
\end{eqnarray}
and we expand $e^{x\bar f_1(s)}$ assuming that $x\bar f_1(s)$ is small. 
From the validity of such an expansion, it is required 
that $|x \bar f_1(s)|\ll 1$. 
As for the case with large $|s|$, 
the integrand becomes exponentially small when $|xs|\gg 1$. 
There we do not have to mind at all even if $x\bar f_1(s)$ becomes 
large negative. 
In the restricted region satisfying $|xs|\alt 1$, 
it is easy to see that $x\bar f_1(s)\ll 1$ is always guaranteed.  
As for the case with small $|s|$, 
we do not have to consider the situation 
in which $|s|$ becomes extremely small 
because there is no requirement on the choice of contour except 
for being inside the saddle points. 
For example, if we choose the contour to be $|s|\agt
1$, $|x \bar f_1(s)| \ll 1$ is guaranteed.
Therefore, 
we find that it is allowed to expand $e^{x\bar f_1(s)}$ 
as $1+x\bar f_1(s)+\cdots$.

After this expansion, introducing a new variable $z$ by 
$e^{-i\pi}z:= sx$, 
the integration along $C_3$ is written as 
\begin{equation}
\psi_{3} (x) \approx 
x^{i \omega \over \kappa} \int_{\bar C_{3}} 
\, dz \, (-z)^{-1 - {i \omega \over \kappa}}
e^{-z}[1+x \bar f_1],
\label{c3}
\end{equation}
where $\bar C_3$ is the contour in the complex $z-$plane 
corresponding to $C_3$. 
Since the integrand becomes
exponentially small at the boundaries, 
we are allowed to continue 
the contour to  $\infty$. 
Then, 
using the integral representation of a gamma function, 
the leading term corresponding to $1$ in the 
square brackets of Eq.~(\ref{c3}) is expressed as   
\begin{equation}
\psi_{3} (x) \approx 
-2 \sinh (\pi \omega / \kappa ) 
\Gamma (-i \omega / \kappa) \, x^{i \omega / \kappa}.
\label{gamma}
\end{equation}

Next, we consider the remaining terms in Eq.~(\ref{c3}).  
Let us express $x\bar f_1$ as $x\bar f_1=\sum a_n (-z)^n$, where
the coefficients $a_n$ are non-dimensional constants. 
Then, by using the integral representation of the gamma function, 
we can evaluate the contribution from each term as 
\begin{equation}
a_n \int_{\bar C_{3}} \, dz \, (-z)^{n-1-i\omega/\kappa} e^{-z}
=-2 a_n \sinh\left(\pi\left({\omega\over\kappa}+in\right)\right)
 \Gamma\left(n-{i\omega\over \kappa}\right),
\end{equation}
and we find that its relative order 
is simply determined by the order of $a_n$. 
Hence, to find the expression correct up to $O(\epsilon^1, \delta^1)$, 
the only term that we must keep is 
\begin{equation}
x \bar f_1(s)\approx-{i\omega \over {2\kappa}}
\left({i\omega \over {\kappa}}+1 \right) \tilde \delta_1(sx)^{-1}.
\end{equation}
Thus, we finally obtain 
\begin{equation}
\psi_{3} (x) = -2 \sinh (\pi \omega / \kappa ) 
\Gamma (-i \omega / \kappa) \, x^{i \omega / \kappa}
\left(1-{i\omega \over {2\kappa}} \tilde \delta_1 
+O(\epsilon^{2}, \delta^{2}) \right).
\label{near3}
\end{equation}

In this section, we approximately solved 
Eq.~(\ref{horizon feq}) with the boundary condition that 
the wave function 
decreases exponentially inside the horizon. 
We evaluated the explicit form of the approximate 
solution in the region $x_1>x>x_2$ as 
\begin{equation}
\psi(x) = \psi_{1}(x) + \psi_{2}(x) + \psi_{3}(x),
\label{x>0}
\end{equation}
where each component $\psi_i(x)$ is 
given by Eq.~(\ref{near12}) or Eq.~(\ref{near3}).  
This expression is correct up to 
$O(\epsilon^1, \delta^1)$ . 

\subsection{the case far from the horizon}

In the region far from the horizon, 
spacetime will become almost flat. 
In this region we assume that the rate of change of $1/\Omega(x)$ 
and $v(x)$ is sufficiently small. 
As we have seen for the asymptotic form of 
solutions in Sec.2, 
we have four independent solutions since the ODE~(\ref{ode}) 
is of fourth order. 
For the solutions with short wavelength  
corresponding to $k_{\pm}$, 
we can use WKB approximation to solve 
Eq.~(\ref{feq}).
On the other hand, for the solutions with long wavelength 
corresponding to $k_{\pm s}$, we can solve Eq.~(\ref{feq}) 
perturbatively by treating the correction 
due to the modification of dispersion relation as small. 
To be strict, we restrict our consideration to 
the region $x>x_2$, where $x_2$ is that given in Eq.~(\ref{x_1}). 
In this region, we assume that the following relations 
\begin{equation}
\Omega{d\over dx} {1\over \Omega},~
{1 \over v}{d\over dx} v,~ 
{1\over (1-v^2)}{d\over dx}(1-v^2) \alt {\omega\over 1-v^2}, 
\label{wkbyuukou}
\end{equation}
are satisfied. 
As for higher order differentiations, 
we also assume that they are all restricted like 
\begin{eqnarray*}
 \Omega {d^2\over dx^2}{1\over \Omega} \alt
\left({\omega\over 1-v^2}\right)^2, \quad
 \Omega {d^3\over dx^3}{1\over \Omega} \alt
\left({\omega\over 1-v^2}\right)^3, \cdots. 
\end{eqnarray*}
By substituting the expansion (\ref{vomegat}), 
we find that these conditions are satisfied even in the 
region close to the horizon. 

Here, we define a quantity $\epsilon(x):=
{\omega\over {k_0 \Omega(1-v^2)^{3/2}}}$, which reduces 
to $\tilde\epsilon(x)$ near the horizon. 
It will be natural to assume that $\epsilon(x)$ takes 
its largest value in the region close to the horizon,  
and hence $\epsilon(x)$ is at most of $O(\epsilon^1)$ 
owing to the restriction $x>x_2$. 
In the following,  
we construct approximate solutions valid up to $O(\epsilon^1)$ 
in the sense of $\delta\psi/\psi$. 

We begin with considering 
the solutions with short wavelength. 
Substituting the expression 
\begin{equation}
\psi = e^{i\int^x dx'\, k(x')},
\label{WKB}
\end{equation}
into Eq.~(\ref{ode}), 
we write down the equation for $k(x)$. 
Neglecting the terms on which 
differentiations with respect to $x$ acted more than 
three times, 
\begin{eqnarray}
{\left(1\over{{k_0}{\Omega}}\right)}^{2}k^{4}
-(1-v^{2})k^{2}-2v{\omega}k+{\omega}^{2}
 & \approx & 
 i{d\over{dx}}\left[2{\left(1\over{{k_0}{\Omega}}\right)}^{2}k^{3}
 -(1-v^{2})k-v{\omega}\right]
\nonumber \\
 & &+{3k'{}^2+4kk''\over k_0^2\Omega^2}
+{12kk'\over k_0^2\Omega}\left({d\over dx}{1\over \Omega}\right)
\nonumber \\
& &
+{5 k^2\over 2k_0^2} \left(\left({d\over dx}{1\over \Omega}\right)^2
+{1\over\Omega}\left({d^2\over dx^2}{1\over \Omega}\right)\right),
\label{feq2}
\end{eqnarray}
is obtained, where $~'~$ is used to represent 
a differentiation with respect to $x$. 
Denoting the left hand side of Eq.~(\ref{feq2}) by $F(k)$,
we find that the first term on the right hand side is expressed as 
\begin{eqnarray*}
{i\over 2}{d\over dx} \left({dF(k)\over dk}\right).
\end{eqnarray*}
We denote the remaining terms on the right hand side by $G(k)$.
{}Following the standard prescription of the WKB approximation,
the terms which contain differentiations with 
respect to $x$ is taken to be small. 
Accordingly, we also expand $k(x)$ in accordance with 
the number of differentiation as 
\begin{equation}
k(x)=k^{(0)}(x)+k^{(1)}(x)+k^{(2)}(x)+\cdots.
\label{g}
\end{equation}
After a slightly long but a straight forward calculation, 
we obtain 
\begin{eqnarray}
k_{\pm} ^{(0)} &=&
\pm k_0 \Omega \sqrt{1-v^2}\left(1\pm v\epsilon(x)
 -{1+2v^2\over 2}\epsilon^2(x)+O(\epsilon^3)\right)\cr
 & = & \pm k_0 \Omega \sqrt{1-v^2}+{v \omega \over 1-v^2}
\mp {(1+2v^{2})\omega^{2} \over{2k_0 \Omega (1-v^{2})^{5/2}}}
+\cdots, 
\label{k^0}
\\
k_{\pm}^{(1)} & = &{i\over2}{d\over{dx}}
{\ln}\left(\pm k_0 \Omega (1-v^{2})^{3/2}
(1\pm 4v\epsilon(x)+O(\epsilon^2))
\right)\cr
&=& {i\over2}{d\over{dx}}
{\ln}\left(\pm k_0 \Omega (1-v^{2})^{3/2}+4vw+\cdots\right),
\\
k_{\pm}^{(2)} &=& \pm k_0\Omega\sqrt{1-v^2}\Biggl\{
{\epsilon^2(x)\over 8\omega^2}
\Biggl[41v^2 v'{}^2+(1-v^{2})\left(
 14(vv')'+18 vv'\Omega\left({1\over \Omega}\right)'\right)
\cr
&&\hspace*{3cm}
+(1-v^2)^2\left(4\Omega\left({1\over \Omega}\right)''+
\biggr[\Omega\left({1\over \Omega}\right)'\biggr]^2\right)\Biggr]
+O(\epsilon^3)\Biggr\}.
\end{eqnarray}

Now we turn to the solutions with small absolute values, 
i.e., $k_{\pm s}$.
In this case, we cannot use the WKB approximation 
because the wavelength is not necessarily short 
compared with the typical scale for the background quantities 
to change. 
However, for the model with the standard dispersion relation, 
we have exact solutions for the field equation 
$k_{\pm s}= \pm\omega/(1\pm v)$. 
We can use them as the leading order approximation, 
which is solutions when we neglect the terms related 
to the modification of dispersion relation. 
If we substitute $k_{\pm s}\approx \pm\omega/(1\pm v)$ 
into the neglected terms, we find 
that all of them have relative order 
higher than $\epsilon^2$. 
At first glance, one may think that the terms corresponding 
to the second and third terms in the square brackets in 
Eq.~(\ref{feq2}) give a correction of $O(\epsilon^0)$, but  
they mutually cancel out. 
As a result, 
the equation to determine the correction 
$\delta k_{\pm s}:=k_{\pm s}\mp\omega/(1\pm v)$ 
is obtained as 
\begin{equation}
 -i\partial_x (1-v^2)\delta k_{\pm s}\pm 2\omega \delta k_{\pm s}=
 {1\over 2k_0^2} e^{-i\int^x k_{\pm s}^{(0)}(x') dx'}
 \left[\partial_x^2{1\over \Omega}\partial_x{1\over \Omega}\partial_x
 +\partial_x{1\over \Omega}\partial_x{1\over
 \Omega}\partial_x^2\right]
 e^{i\int^x k_{\pm s}^{(0)}(x') dx'}=:H_{\pm}(x). 
\label{deltak}
\end{equation}
The right hand side consists of the terms 
related to the modification of dispersion relation, 
and they are small of $\omega k_{+s}^{(0)}\times O(\epsilon^2)$. 
Different from the case for short wavelength modes, 
the equation to determine the correction becomes 
a differential equation. 
Therefore, we can say that the correction stays of 
$O(\epsilon^2)$ only when we are interested in 
the behavior of the solution within a small region 
such as $x_1>x>x_2$. 
Once an extended region is concerned, there is no reason 
why the correction stays of $O(\epsilon^2)$. 
In fact, we need to know the behavior of the 
solution both at infinity and in the matching region $x_1>x>x_2$. 
In such a case, the correction much larger than 
$O(\epsilon^2)$ can appear as explained in detail in  
Appendix \ref{appendixC}.  

Nevertheless, the origin of this correction is 
the effect of scattering due to the modified 
dispersion relation. 
Even if the observed spectrum of the emitted 
particles deviates from the thermal one 
due to this effect, it is still possible to adopt 
the interpretation that   
the spectrum is modified by the scattering 
during the propagation to a distant observer 
though it was initially thermal. 
Hence, we think that this effect should be 
discussed separately from the present issue. 

However, to be precise, we consider the case 
that the condition (\ref{wkbyuukou}) replacing $\alt$ 
with $\ll$ is satisfied. 
This is the case when $\omega$ is sufficiently large 
or when the functions $v(x)$ and $1/\Omega(x)$ 
rapidly converge to some constants at $x\agt x_1$. 
In such cases, we can think of 
the first term on the left hand side of Eq.~(\ref{deltak}) 
as small. Then, solving Eq.~(\ref{deltak}) iteratively, 
we find that the correction stays of $O(\epsilon^2)$. 
Therefore, we obtain
\begin{equation}
k_{\pm s} \approx \pm{\omega \over 1\pm v}(1+O(\epsilon^{2})). 
\end{equation}

Consequently, we find that 
the solutions which behave like $e^{ik(x\to\infty) x}$ 
at infinity are given by 
\begin{eqnarray} &&
{\psi}_{\pm}(x) = e^{i\int^x k_{\pm}(y) dy},\quad\quad
\psi_{\pm s}(x) = e^{i\int^x k_{\pm s}(y) dy},
\label{formalwkb}
\end{eqnarray}
where the integral constants are chosen appropriately. 
Here we recall that what we wish to know is not $k(x)$ but 
$\int^x dy\, k(y)$. 
Although we are keeping track of the error in 
the expression of $k(x)$, 
we cannot evaluate the error in 
the integral $\int^x_\infty k(y) dy$ when it is 
integrated from $\infty$ to the matching region where $x_1>x>x_2$.
To overcome this difficulty, we need to make use of  
the existence of a conserved current 
\begin{equation}
j=A(k(x)) e^{-2\int^x k_{(I)}(y) dy},
\end{equation}
where
\begin{eqnarray}
A(k(x))=&&\omega v+(1-v^2)k_{(R)} \cr 
&&-{1\over k_0^2\Omega^2}\left\{2k_{(R)}(k_{(R)}^2-k_{(I)}^2)
 +4k_{(R)}\partial_x k_{(I)}+2k_{(I)}\partial_x k_{(R)}
 -\partial_x^2 k_{(R)}\right\}\cr
&&+{2\over k_0^2}\left({1\over\Omega}\left({1\over\Omega}\right)'\right)
 \left(-2k_{(R)}k_{(I)}+\partial_x k_{(R)}\right)
+{1\over 2k_0^2}\left({1\over\Omega}\left({1\over\Omega}\right)'\right)'
  k_{(R)}, 
\end{eqnarray}
and $k_{(R)}$ and $k_{(I)}$ are the real and the imaginary parts 
of $k$, respectively. 
The derivation of $j$ is given in Appendix \ref{appendixD}. 

We evaluate this conserved current $j$ at $x\to\infty$, 
where all terms that contain differentiations with respect 
to $x$ vanish there.  
Adopting the normalization $\vert\phi\vert^2 = 1$ at $x\to\infty$,  
$j$ is determined as
\begin{equation}
 j=\omega v_{\infty}+(1-v_{\infty}^2)k_{\infty}
   -{2\over k_0^2\Omega_{\infty}^2} k_{\infty}^3. 
\end{equation}
For $l=\pm,\pm s$, by substituting $k_{l\infty}$ into 
the expression of $j$ in place of $k_{\infty}$, 
we also define the conserved current $j_l$ 
corresponding to $k_l$. 

Owing to the conservation of $j$, 
\begin{equation}
 \psi_l(x)=\sqrt{j_l\over A(k_l(x))} 
  e^{i\int^x k_{l(R)}(y) dy}.
\label{hozonpsi}
\end{equation}
By using this improved expression, 
we can calculate the explicit form of $\psi_l(x)$ 
in the region $x_1>x>x_2$ without any ambiguity 
except for the constant phase factor that does not 
alter the absolute magnitude of the wave function.   
Expanding the expression (\ref{hozonpsi}) in powers of 
$\tilde \delta_i$ with the substitution of Eqs.~(\ref{vomegat}), 
we evaluate $\psi_l(x)$ keeping the terms up to $O(\delta^1)$.
Here in evaluating $A(x)$, 
the terms that include $k_{(I)}$ becomes higher order 
in $\epsilon$, and hence we can neglect them all. 
Consequently, we obtain
\begin{eqnarray}
&&
{\psi_{\pm}(x)\over \sqrt{j_{\pm}}}
 ={e^{i\alpha_{\pm}}\over {\sqrt{k_0 \Omega_0}}}
(2 \kappa x)^{-{3\over 4}-{i\omega \over {2\kappa}}}
\exp  \left( \pm i {2\over 3} k_0 \Omega_0 \sqrt{2 \kappa} x^{3\over 2} 
+W_{\pm}+O(\epsilon^{2}, \delta^{2})\right),
\cr &&
{\psi_{+s}(x)\over\sqrt{j_{+ s}}} = e^{i\alpha_{+s}}
x^{i\omega \over \kappa}
\exp\left(-{i\omega \over {2 \kappa}} \tilde \delta_1 
  +O(\epsilon^2,\delta^2)\right),
\label{fwkb}
\end{eqnarray}
where $W_+$ and $W_-$ are no different from $W_2$ and $W_1$ 
in Eq.~(\ref{name3}), respectively.
As noted above, there appears an integration constant $\alpha_l$ 
that cannot be determined by the present analysis, 
but it is guaranteed to be a real number.
Here, we did not give the explicit form of $\psi_{-s}$ 
because we do not use it later.  

By comparing (\ref{x>0}) with (\ref{fwkb}), 
we find that the solution (\ref{x>0}) obtained in the region 
close to the horizon is matched to the solutions obtained 
in the outer region like 
\begin{equation}
\psi(x)=\sqrt{2 \pi \kappa}
 \left(
  e^{-{\pi \omega  
  \over{2 \kappa}}+i\alpha'_-} {\psi_-\over \sqrt{j_-}}+ 
  e^{{\pi \omega \over 2 \kappa}+i\alpha'_+}
 {\psi_+\over \sqrt{j_+}} 
  \right) 
- 2 \sinh \left( {\pi \omega \over \kappa} \right)
\Gamma \left(-{i\omega \over \kappa} \right) 
 e^{i\alpha_{+s}}{\psi_{+s}\over \sqrt{j_{+s}}},
\label{fsolution}
\end{equation}

where $\alpha'_{\pm}$ are also real constants.
From this expression, we can read the coefficients $c_{+s},c_{-}$ 
as 
\begin{eqnarray}
&&
c_{+s}=- 2 \sinh \left( {\pi \omega \over \kappa} \right)
\Gamma \left(-{i\omega \over \kappa} \right)
 {e^{i\alpha_{+s}}\over \sqrt{j_{+s}}}(1+O(\epsilon^2,\delta^2))
\cr &&
c_{-}=\sqrt{2 \pi \kappa}e^{-{\pi \omega \over{2 \kappa}}} 
{e^{i\alpha'_{-}}\over \sqrt{j_{-}}}
\left( 1 - {4v\omega\over{k_0\Omega}}(1-v^{2})^{-3/2}\right)^{-1/2}
(1+O(\epsilon^2,\delta^2)).
\end{eqnarray}
The factor $\omega' v_g$ appears in the formula (\ref{number}) for 
the expectation value of the created particles.
By differentiating 
the dispersion relation at infinity (\ref{disp}) with
respect to $k_{\infty}$, 
this factor is easily calculated as  
\begin{equation}
 \omega'(k_{\infty}) v_g(k_{\infty})
 :=(\omega-v k_{\infty}) {d\omega(k_\infty)\over dk_{\infty}}
  =v_{\infty}\omega +(1-v_{\infty}^2)k_{\infty}
  -{2k_{\infty}^3\over k_0^2\Omega_{\infty}^2},   
\end{equation}
and we find it to coincide with the conserved current. 
Thus, by considering the combination of 
$\omega'(k_{+s}) v_g(k_{+s})\vert c_{+s}\vert^2$
the factor $j_{+s}$ in $c_{+s}$ cancels, and 
the same is also true for $k_-$. 
Finally the expectation value of the number of created particle is
evaluated as
\begin{equation}
N(\omega)={ 1 \over {e^{2 \pi \omega / \kappa}-1}}
(1+O(\epsilon^{2},\delta^{2})).
\end{equation}

\section{analytic and numerical studies of 
the deviation from hawking spectrum}

\vspace*{5mm}
{\renewcommand{\arraystretch}{1.6}
\begin{center}

\begin{tabular}{c||c|c|c|c}
$~~~~~~$ & $~~\epsilon^{-1}~~$ & $~~\epsilon^{0}~~$ 
& $~~\epsilon^{1}~~$ & $~~\epsilon^{2}~~$
\\
\hline
&&&&\\[-6.1mm]
\hline
{$\delta^{0}$} & $x^{3/2}$ & $x^0$ & $x^{-3/2}$ & $x^{-3} $
\\
\hline
$\delta^{1}$ & $x^{5/2}$ & $x^{1}$ & $x^{-1/2}$ & $x^{-2}$ 
\\
\hline
$\delta^{2}$ & $x^{7/2}$ & $x^{2}$ & $x^{1/2}$ & $x^{-1}$ 
\\
\hline
$\delta^{3}$ & $x^{9/2}$ & $x^{3}$ & $x^{3/2}$ & $x^{0}$ 
\end{tabular}
\vspace*{5mm}

{\small TABLE I. 
Table to explain the matching procedure.}
\end{center}
}

In the preceding section, to obtain the flux of the created 
particles observed in the 
asymptotic region, where $v(x)$ is essentially constant, 
we propagated the near-horizon solution (\ref{x>0}), which 
satisfies the appropriate boundary condition, to the 
infinity by matching it with the outer-region solutions 
(\ref{fwkb}), which are valid in the region 
distant from the horizon. As a result, we could determine 
the coefficients $c_{l}$ and we found that the thermal 
spectrum is reproduced up to $O(\epsilon^1, \delta^1)$.

To explain the matching procedure in more detail, 
here we present a table. 
In the construction of the near-horizon solution, 
the equation to be solved was expanded with respect to 
$\delta$, and we obtained an equation which 
correctly determines the terms up to $O(\delta^1)$. 
These terms correspond to the first two lines in the above table. 
Then, expanding them with respect to $\epsilon$ 
by restricting our consideration to the region $x_2<x<x_1$, 
we obtained the expression (\ref{x>0}) with 
(\ref{near12}) and (\ref{near3}), which contains the  
terms corresponding to the first $2\times 3$ elements 
in the table. 

On the other hand, 
in the region distant from the horizon, 
we first considered an expansion of 
the solution with respect to $\epsilon$, and 
calculated the corrections up to $O(\epsilon^1)$.
Namely, the terms corresponding to the first three columns in the above 
table are obtained. 
As the next step, in the region of $x_2<x<x_1$, 
we expanded this expression also with respect to $\delta$ up to 
$O(\delta^1)$ by substituting (\ref{vomegat}) into Eq.~(\ref{name3}), 
and we obtained the first $2\times 3$ elements in the table. 
The expressions that we finally obtained 
are the outer-region solutions~(\ref{fwkb}). 

Such two fold expansions in both schemes are simultaneously 
valid only in the region 
$x_1<x<x_2$, where both $\tilde \epsilon(x)$ and $\tilde \delta_i(x)$ 
are small. 
As mentioned above, as long as $k_0$ is taken to be sufficiently 
large, this overlapping region always exists. 
Since both expressions obtained by using the above two 
different schemes are approximate solutions of the same equation, 
they must be identical if we take an appropriate 
superposition of the four independent solutions. 
In fact, we found that the near-horizon solution (\ref{x>0}) can be 
written as a superposition of the outer-region solutions 
as given in Eq.~(\ref{fsolution}).
Now, let us look at the above table again. 
For each element in the table, 
we have assigned a power of $x$ that the 
corresponding terms possess. As we mentioned above, 
we need to choose an appropriate 
superposition of the four independent outer-region 
solutions to achieve a successful matching. 
The coefficients which determine the weight of this 
superposition is nothing but $c_l$. 
Now we should note that 
the condition to determine the coefficients $c_l$ 
will be completely supplied by matching the  
$x$-independent elements. Once these coefficients are 
determined, the agreement of the other $x$-dependent 
terms must be automatic from the consistency. 
The leading order $x$-independent elements 
consists of the terms of $O(\epsilon^0\delta^0)$, 
and it is easy to see that the second lowest one 
consists of the terms of $O(\epsilon^{2} \delta^{3})$. 
This fact tells us that the possible modification  
of the coefficients $c_{+s}$ and $c_-$ is at most 
of $O(\epsilon^{2} \delta^{3})$, and hence 
the possible deviation from the thermal radiation 
starts only from this order. 
Thus we find 
\begin{equation}
N(\omega)={ 1 \over {e^{2 \pi \omega / \kappa}-1}}
(1+O(\epsilon^{2} \delta^{3})).
\label{thermal}
\end{equation}

Next we investigate the deviation from thermal radiation 
$\delta N/N:=(N-N_{thermal})/N_{thermal}$ in more detail. 
There are three different quantities of $O(\delta)$ as given in 
(\ref{ddelta}). 
We write them down as 
\begin{equation}
 \tilde\delta_0=x v'|_{x=0} =\kappa x, 
 \quad \tilde\delta_1=x \kappa^{-1} v''|_{x=0} =:(\kappa x)b_1, 
 \quad \tilde\delta_2=x \Omega_0 \left({1\over \Omega}\right)'_{x=0} 
             =:(\kappa x)b_2,
\end{equation}
where we introduced 
non-dimensional model parameters $b_1$ and $b_2$. 
We also have various quantities of $O(\delta^2)$ 
and of $O(\delta^3)$ consisted of higher derivatives of 
$v$ and $1/\Omega$. They are  
\begin{equation}
 x^2 \kappa^{-1} v'''|_{x=0}
             =:(\kappa x)^2 b_3,
 \quad\quad x^2 \Omega_0 \left({1\over \Omega}\right)'_{x=0}
             =:(\kappa x)^2 b_4,
\end{equation}
and
\begin{equation}
 x^3 \kappa^{-1} v^{(4)}|_{x=0}
             =:(\kappa x)^3 b_5,
 \quad\quad x^3 \Omega_0 \left({1\over \Omega}\right)''_{x=0}
             =:(\kappa x)^3 b_6,
\end{equation}
respectively. Also, $b_3$, $b_4$, $b_5$ and $b_6$ are non-dimensional 
model parameters.  
One may suspect that terms including factors 
proportional to $\kappa^{-2}$ such as $x \kappa^{-2} v'''|_{x=0}$ 
might appear 
among the correction terms of $O(\epsilon^2\delta^3)$.
However, by repeating the same calculation that was given in Sec.4 
with extra higher order derivative terms, 
we can verify that such factors do not appear.   
From this notion, we can expect that the deviation from  
the thermal spectrum is given by 
\begin{eqnarray}
 {\delta N\over N} ={\kappa^2\over k_0^2\Omega_0^2}\Biggl[
                &&  \left(a_{000}+a_{001} b_1+a_{002} b_2
                          +(\mbox{7 other terms})\right) \cr
                &&  +\left(a_{03}+
                        a_{13} b_1+
                        a_{23} b_2\right) b_{3}
                    +\left(a_{04}+
                        a_{14} b_1+
                        a_{24} b_2\right) b_{4}
                   +a_{5} b_{5}
                   +a_{6} b_{6}\Biggr]
                   +O(\epsilon^4\delta^{6}),   
\label{dev}
\end{eqnarray}
where ``$a$''s are some functions of $\omega/\kappa$ which are 
independent of model parameters. 

Now, we numerically confirm that the deviation actually 
starts from this order. The following results are 
obtained by using {\it MATHEMATICA}. 
As an example, let us consider a model given by 
\begin{eqnarray}
 v & = & -{1\over 2}e^{-2 x - 3 x^2} 
   -{1\over 2}, \label{vmodel} \\
\Omega & = & 9 e^{-x-x^2/2}+1. 
\label{Omegamodel} 
\end{eqnarray}
For this model, we have $\kappa=1$, $b_1=1$ and $b_2=9/10$, 
and the other parameters also do not vanish.  
For this fixed model, we numerically calculated the deviation 
from the thermal spectrum for various values of $1/k_0^{2}$. 
The frequency $\omega$ was fixed to 1 since our main interest 
is in the modes whose observed frequency at infinity becomes
comparable with Hawking temperature ($= \kappa /2 \pi$). 
The results of the numerical calculation 
are shown in Fig.3 by the filed circles. 
The horizontal axis is $\log (1/k_0)$ and the vertical one 
is $\log \delta N/N$. 
The data points are fitted well by a liner function (the solid line)
with its gradient, $1.99742$, 
which perfectly agrees with  
the expectation represented by Eq.(\ref{dev}). 

At this point, one may notice that 
the deviation we obtained here is much larger 
than that given by Corley and Jacobson\cite{JacCor}, 
in which a model with   
\begin{equation}
 v = {1\over 2} \{\tanh [(2\kappa x)^{2}]\}^{1/2}-1,
\end{equation}
and $\Omega\equiv 1$ was considered
\footnote{T. Jacobson suggested us 
the existence of this discrepancy.}.
The outstanding feature of their model is that 
$b_1=b_2=b_3=b_4=b_5=b_6=0$. Hence, the terms 
of $O(\epsilon^2\delta^3)$ in Eq.(\ref{dev}) reduce to  
$a_{000}{\kappa^2 /k_0^2\Omega_0^2}$. 
If $a_{000}\equiv 0$, all the terms of $O(\epsilon^2\delta^3)$ in 
the deviation $\delta N/N$ disappear, and 
it turns out to be $O(\epsilon^4\delta^6)$. 
If so, the discrepancy between two calculations can be understood. 
To show that this is certainly the case, we repeated the  
numerical calculation for the same model that was discussed in 
Ref.\cite{JacCor}. The resulting $\delta N/N$ calculated for various 
values of $1/k_0$ were also plotted in Fig.3 by the 
open squares. 
Again, the data points in logarithmic plot are 
fitted well by a linear function. But, this time, 
its gradient is 4.06935, which indicates that the deviation 
is actually caused by the terms of $O(\epsilon^4\delta^6)$.

Now, we can conclude that $a_{000}\equiv 0$. 
Although this result might be interesting, 
we do not pursue this direction of study in this paper.  
Here, we would like to focus on another interesting 
aspect that is anticipated by the expression (\ref{dev}). 
With moderate values of model parameters, 
the deviation $\delta N/N$ 
stays small for a sufficiently large $k_0$.  
However, conversely, we can expect that the deviation 
from the thermal spectrum becomes large 
if we consider some extreme modifications 
of the special reference frame.  
Especially, when we consider the limiting case in which  
$\Omega_0^2\to 0$, the expression (\ref{dev}) diverges. 
Although the approximation used to obtain the analytic expression (\ref{dev}) 
is no longer valid in this limit, we can still expect 
that the resulting spectrum will significantly differ 
from the thermal one. 
As we mentioned below Eq.(\ref{newv}), there is a 
lower bound on $\Omega^2(x)$. The possible smallest value 
of $\Omega^2(x)$ is $1-\tilde v^2$, which is realized when $v(x)=0$. 
Hence, we find $v(x)\approx 0$ near $x=0$ in this limiting case. 
Recall that $v(x)$ was the coordinate 
velocity of the integration curves of $u$. 
Hence, vanishing $v(x)$ means that we adopt
the reference frame corresponding to the static observers. 

Here, we present the results of our numerical calculation, 
which shows that the deviation $\delta N/N$ can be large 
for some cases.   
Since we also want to demonstrate that the drastic 
change of spectrum can occur just in the consequence 
of the change of the special reference frame, 
we varies only the function $\gamma(\tilde x)$, which was defined
in Eq.(\ref{tconst}).  
The model of $\tilde v(\tilde x)$ is kept unchanged.  
As for a model of $\tilde v(\tilde x)$, we assume the 
same form that is given in Eq.(\ref{vmodel}). 
As for $\gamma(\tilde x)$, we adopt
\begin{equation}
 \gamma={-\tilde v-(1/2)\over \rho^{-1} -\tilde v}, 
\end{equation}
with $\rho \in [0,1)$.
$\rho=0$ corresponds to the original model associated with 
the freely falling observers, and 
$\rho=1$ corresponds to the case with $\Omega_0^2=0$. 
With this choice of $\gamma(\tilde x)$, the following two conditions 
are satisfied.  
One is that $v(x)$ stays negative for all positive $x$. 
The other is that $\Omega_{\infty}^2= 1$. 
We calculated the deviation $\delta N/N$ 
for various values of $\rho$, and the results were shown in 
Fig.4. As was expected, the deviation becomes large 
for small $\Omega_0^2$. This plot raises an interesting speculation   
that $N$ might converge to $0$ in the $\Omega_0^2\to 0$ limit.
Although we have not confirmed it yet, it is very likely that 
this is the case because the situation in this limit is very 
similar to the case in which we set a static mirror surrounding 
the event horizon of the black hole. 
The calculation for small $\Omega_0^{2}$ were tried, 
but it was found to be out of validity of our present 
computation code.
Anyway, we conclude that, even if $\kappa/k_0$ is sufficiently small, 
the deviation from the thermal spectrum can be large 
if the combination $\kappa/k_0\Omega_0$ becomes large. 
It should be noted that the integration curves of $u$ 
are not required to be suffered from infinite acceleration 
to achieve such a small but non-zero $\Omega_0^2$.

\begin{figure}[tbh]
\vspace*{0cm}
\centerline{\epsfxsize7cm \epsfbox{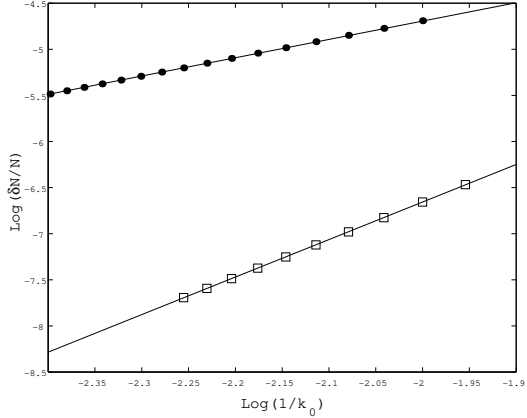}}
\caption{The logarithmic plot of $\delta N/N$ as a function 
of $1/k_0$ for two different choice of $v(x)$. 
The filled circles(our model) and the open squares(model in 
Ref.[5]) represent the 
numerical data points for respective models. 
Each solid line corresponds to a linear function
which fits the data points. 
}
\label{fig4}
\end{figure}

\begin{figure}[tbh]
\vspace*{0cm}
\centerline{\epsfxsize7cm \epsfbox{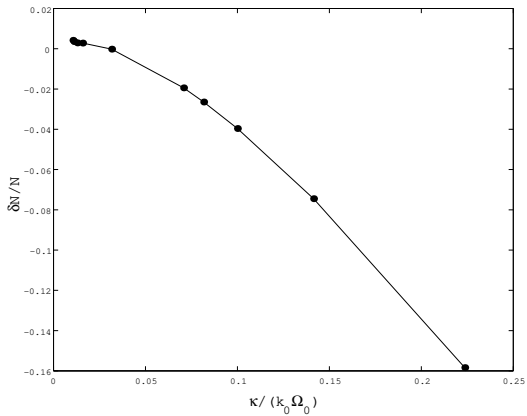}}
\caption{A plot of $\delta N/N$ as a function of $1/\Omega_0$ for
the model given by Eq.(5.6).
}
\label{fig5}
\end{figure}

\section{Conclusion}
We studied the particle creation in a model which 
is a generalization of the Unruh's toy model. In his model, 
the field equation for a scalar field is modified by introducing 
a non-standard dispersion relation. To do so, we necessarily 
violates the Lorentz invariance. This radical change of theory 
is originally motivated by the possible existence of the effect 
due to the unknown physics at Planck scale. 
However, as is explained in Introduction, 
there is another point of view, on which  
it is also meaningful 
to study this model as an effective theory which takes into 
account the interaction between various fields 
even if we believe that the Lorentz invariance is exact,  

In the original model, the dispersion relation is modified 
on the basis of the freely falling observers. 
In our present work, we generalized the choice of the 
reference frame with respect to which we set the non-standard 
dispersion relation. 
Extending the analytic method developed by Corley~\cite{Corley},
we have shown that the thermal spectrum of radiation from a black 
hole is almost reproduced as long as the 
modification of the special reference frame is not too extreme. 
In this analysis, we assumed $\omega\approx \kappa$, where   
$\omega$ is the frequency of the emitted photon observed at 
the spatial infinity and  
$\kappa$ is the surface gravity of the black hole. 
We have also obtained a strong suggesting that the deviation 
from the exact thermal spectrum appears from 
$O(\kappa^2/k_0^2)$, where 
$k_0$ is the typical wave number corresponding to 
the modification of the dispersion relation. 
This speculation has been confirmed numerically.

Of course, we should not stress this small 
deviation from the Hawking spectrum. 
In the ordinary model with the Lorentz invariant 
dispersion relation, 
the thermal radiation at the temperature 
$T$ for the static observer is observed as 
the thermal radiation at the temperature 
$\sqrt{1-\beta\over 1+\beta} T$ for the observer moving 
with the radial velocity $\beta$. 
This argument holds in general whatever the source of the 
outward pointing radiation is 
because it is a direct consequence of $\omega=k$. 
However, in the present modified model, 
the Lorentz invariance is violated from the beginning. 
We can easily see that $(\omega-k)/\omega$ is also of 
$O(\omega^2/k_0^2)$. Hence, even if the exact Hawking spectrum 
is reproduced for one specific free-falling observer, 
it cannot be so for the other free-falling observers. 

On the other hand, the result that we obtained analytically 
also suggests that the deviation from the thermal spectrum 
can be large if we consider some extreme situations. 
With the aid of numerical methods, 
we also examined one of such extreme situations.
We considered a sequence of 
different special reference frames which ranges from the 
case in which the observers associated with the 
special reference frame are freely falling into black hole  
to the case in which they are kept from falling into it. 
We found that, in the latter limiting case, 
the spectrum of radiation can significantly 
differ from the thermal one, 
even though $\omega^2/k_0^2$ is small. 
It will be important 
to study the physical meaning of this result. 
But, since the central issue of this paper is 
to develop the analytic treatment of our new model,  
we have not performed detailed numerical studies yet.   
We will return to this issue in future publications.

\acknowledgements
{We would like to thank Misao Sasaki for useful suggestions and 
comments. Y.H. also thanks Fumio Takahara for his continuous 
encouragement and Uchida Gen for the useful 
conversation with him. Additionally, he thanks Shiho 
Kobayashi for his appropriate advice. Lastly we wish to thank Ted
Jacobson for fruitful discussions in YKIS'99. 
}

\appendix

\section{the contour modified for the correction term}
\label{appendixA}

\begin{figure}[tbh]
\vspace*{0cm}
\centerline{\epsfxsize5cm \epsfbox{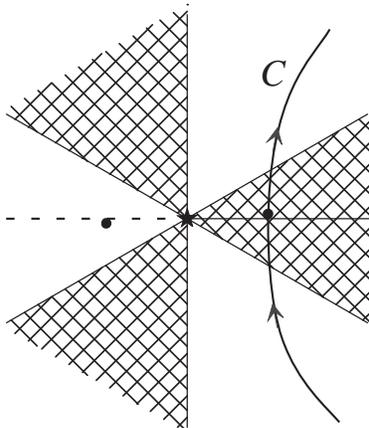}}
\caption{The integration contour satisfying the boundary condition that 
the wave function decays exponentially fast inside the 
event horizon when the terms of higher order 
in $\delta$ can be neglected. In the directions indicated 
by hatched regions, $\hat\psi(s)$ increases exponentially. 
Filled circles represent the saddle points $s_{\pm}$ and 
the dashed line represents the branch cut.  
}
\label{fig2}
\end{figure}

In this appendix, we explain how we choose 
the contour of integration $C$ in Eq.~(\ref{int2}) 
in more detail.
As shown in Ref.~\cite{Corley}, the contour $C$ 
in Fig. \ref{fig2} satisfies the condition 
that the wave function exponentially decreases inside 
the horizon 
when $W^{(1)}$ and higher order corrections are neglected.  
When $W^{(1)}$ are taken into account, however,   
the contour of integration needs to be modified. 
The correction $W^{(1)}$, 
contains the terms proportional to $s^{5}$, while 
the terms of the highest power in the main component $W^{(0)}$ 
is proportional to $s^{3}$. 
As a result, $|W^{(1)}|$ becomes larger than $|W^{(0)}|$ 
when $|s|$ becomes large.
Hence, from the validity of approximation, the contour of
integration must be modified to be contained in the region 
that satisfies the condition $|W^{(0)}| \gg |W^{(1)}|$.
By comparing the absolute value of the $s^{3}$ term in $W^{(0)}$ with 
that of the $s^{5}$ terms in $W^{(1)}$,
the allowed region for the contour to move is found to be restricted by 
\begin{equation}
|s|< s_{max} = \mbox{min} 
\left( k_0\Omega_0, k_0\Omega_0\sqrt{\kappa^{2} 
\over{\kappa_1^{2}}} , k_0\Omega_0 
\sqrt{\kappa\Omega_1\over \Omega_0} \right),
\label{region1}
\end{equation}
Thus, we modify the contour 
not to run into infinity but to terminate at points contained 
in the region $|s|<s_{max}$ as shown in Fig.~\ref{fig1}.  

Because of this modification of the contour of integration, 
the boundary terms in Eq.~(\ref{fourier}) no longer vanish. 
However, since $\hat \psi (s_{max})$ is exponentially small at 
both end points, 
we can expect that the correction due to 
the boundary terms is negligiblly small.

\section{evaluation of integration about saddle points}
\label{appendixB}

In this appendix, we explain the details 
how to evaluate Eq.~(\ref{anten3}). 
We first consider the exponent $xf(s_+)$.  
We evaluate it as an expansion around $s=s_0$ like 
\begin{eqnarray}
xf(s_+) = &&
xf_0(s_0) + f'_0(s_0) (xs_1) 
 +{1 \over 2x} f''_0(s_0) (xs_1)^{2}+\cdots 
\cr &&
 +xf_1(s_0) + f'_1(s_0) (xs_1) 
 +{1 \over 2x} f''_1(s_0) (xs_1)^{2} + \cdots.  
\end{eqnarray}
The first term in the first line of the right hand side 
is zeroth order in $\delta$, and the second term vanishes 
identically. 
The other terms in the first line  
are quadratic or higher order in $\delta$. 
The terms in the second line are proportional to 
$\delta^1, \delta^2, \delta^3, \cdots$, respectively.   
Thus we find 
\begin{equation}
 xf(s_+)= xf_{0}(s_{0})+xf_{1}(s_{0})+O(\delta^{2}). 
\label{part1}
\end{equation}
Using Eqs.~(\ref{xf0}), (\ref{xf1}) and (\ref{sai}), 
$xf(s_+)$ is found to be given by 
\begin{eqnarray}
xf(s_{+}) &=& -\left(\sqrt{{-\tilde\epsilon^{2}\over 2}}\right)^{-1}
 \left({2\over 3} -{1\over{10}}\tilde\delta_{0} 
 +{1\over{10}}\tilde\delta_{1}-{2\over5}\tilde\delta_{2}\right)
\cr &&
+{\tilde\delta_{0}\over4}\left(3+i{\omega \over {\kappa}}\right)
-{\tilde\delta_{1}\over4}\left(3-i{\omega \over {\kappa}}\right)
+2\tilde\delta_2
-\left(1+i{\omega \over {\kappa}}\right)
\log (k_0 \Omega_0 \sqrt{-2\kappa x})
\cr &&
+{1\over4}\left(1+i{\omega \over {\kappa}}\right)
\sqrt{{-\tilde\epsilon^{2}\over 2}}
\left[-\left(1+i{\omega \over {\kappa}}\right)
+{\tilde\delta_0 \over 4}\left(9+i{\omega \over {\kappa}}\right)
-{\tilde\delta_1 \over 4}\left(9-15i{\omega \over {\kappa}}\right)
+\tilde\delta_2\left(5-3i{\omega \over {\kappa}}\right)\right] 
\cr
&&+O(\delta^{2}, \epsilon^{2}).
\end{eqnarray}

Next,  we evaluate $f''(s_+)$. 
Again we expand it around $s=s_0$ as 
\begin{eqnarray}
{1 \over x }f^{''}(s_+) = &&
{1 \over x }f''_0(s_0)+{1 \over x^2}f'''_0(s_0) (xs_1) +\cdots
\cr &&
+{1 \over x }f''_1(s_0)+{1 \over x^2}f'''_1(s_0) (xs_1)
+ \cdots.  
\end{eqnarray}
The power indices with respect to $\delta$ 
of the respective terms in the first line on 
the right hand side are $0,  1,  \cdots$.  
Those in the second line 
are $1,  2,  \cdots$.  
Hence, the expression for $f^{''}(s_+)/x $ 
which is correct up to $O(\delta)$ is given by 
\begin{equation}
 {1\over x} f''(s_+)\approx 
{1\over x} f''_0(s_0)+
{1\over x^{2}} f'''_0(s_0) (x s_1)+{1\over x} f''_1(s_0).
\label{ddxf}
\end{equation} 
As for this factor, it is not necessary to find the second 
order correction in $\epsilon$. 
Hence, we can use the truncated expressions for $x s_0$ and 
$x s_1$ obtained by discarding 
the terms of $O(\epsilon)$ in Eqs.~(\ref{sai}) and  
(\ref{anten2p}). 
Substituting these into (\ref{ddxf}), we find 
\begin{equation}
{1\over x} f''(s_+) = 2\sqrt{{-\tilde\epsilon^2\over{2}}}\left[
1+{3\over 4}\tilde\delta_0 -{3\over 4}\tilde\delta_1
+3\tilde\delta_2 +
\sqrt{{-\tilde\epsilon^2\over{2}}}
 \left(1+i{\omega \over{\kappa}}\right)
 \left(1+{3\over 2}\tilde\delta_0 -{3\over 2}\tilde\delta_1 
  +6\tilde\delta_2\right)
 +O(\epsilon^2, \delta^2)\right].
\end{equation}

{}Finally, we evaluate the second and third terms 
in the square brackets on the right hand side of Eq.~(\ref{anten3}).
As before, we write $f'''(s_{+})/x^{2}$ as
\begin{eqnarray}
{1\over x^{2}} f'''(s_+) &=& {1\over x^{2}}f'''_{0}(s_{0}) 
+{1\over x^{3}}f^{(4)}_{0}(s_{0})(xs_1)+\cdots 
\cr &&
{1\over x^{2}}f'''_{1}(s_{0}) + {1\over x^{3}}f^{(4)}_{1}(s_{0})
(xs_1)+\cdots .
\label{fd3}
\end{eqnarray}
We evaluate the order of each term on the right hand side in
this equation. 
Then, we find that the respective terms in the first line are of 
$O(\epsilon^2), O(\epsilon^{3}\delta), O(\epsilon^{3}\delta^{2}), 
O(\epsilon^{3}\delta^{3}), \cdots$. 
Those in the second line are of $O(\epsilon^{2}\delta), 
O(\epsilon^{2}\delta^2), O(\epsilon^{2}\delta^3), \cdots$. 
One may notice that the order in the first line does not 
change regularly. This is because 
the second term in the last line in Eq.~(\ref{xf0}) vanishes 
if it is differentiated more than four time. 
Since the leading term in $f^{''}(s_+)/x$ is $O(\epsilon^1)$,
we can neglect the terms of $O(\epsilon^3)$ or higher in Eq.~(\ref{fd3}).
Furthermore, we do not have to keep the terms of $O(\delta^2)$ 
or higher. 
Therefore, we have only to retain the terms 
$f'''_{0}(s_{0})/x^{2}$ and $f'''_{1}(s_{0})/x^{2}$. 
Substituting $x s_{0+}\approx -(-\tilde\epsilon^{2}/2)^{-1/2}$ into these 
two terms, $(xf'''(s_+))^2 / (xf''(s_+))^{3}$ is evaluated as 
\begin{equation}
{(xf'''(s_{+}))^{2}\over (xf''(s_{+}))^{3}}=
{1\over2}{\sqrt{-\tilde\epsilon^{2}\over 2}}
\left(1+{15\over4}\tilde\delta_0 
-{15\over4}\tilde\delta_1+15\tilde\delta_2 \right)
 +O(\epsilon^{2}, \delta^{2}).
\end{equation}

As for $xf^{(4)}(s_+) / (xf''(s_+))^{2}$, 
similarly we have 
\begin{eqnarray}
 {1\over x^{3}} f^{(4)}(s_+) &=& {1\over x^{3}}f^{(4)}_{0}(s_{0}) 
 +{1\over x^{4}}f^{(5)}_{0}(s_{0})(xs_1)+\cdots 
 \cr &&
 {1\over x^{(3)}}f^{(4)}_{1}(s_{0}) 
 + {1\over x^{4}}f^{(5)}_{1}(s_{0})(xs_1)+\cdots .
\end{eqnarray}
The order of respective terms in the first line are 
$O(\epsilon^{4}), O(\epsilon^{4}\delta), \cdots$, 
and those in the second line are
$O(\epsilon^{3}\delta), O(\epsilon^{3}\delta^{2}), \cdots$.
This time, only the term that we must keep is 
$f_{1}^{(4)}(s_{0})/x^{3}$. 
Therefore, we find 
\begin{equation}
{xf^{(4)}(s_+)\over{(xf''(s_+))^2}}
=3\sqrt{{-\tilde\epsilon^{2}\over 2}}
(\tilde\delta_0 - \tilde\delta_1 +4\tilde\delta_2)
+O(\epsilon^{2}, \delta^{2}).
\end{equation}

Substituting all the above results into Eq.~(\ref{anten3}),
finally we obtain Eq.~(\ref{fsolution}).

\section{wave propagation in the modified model}
\label{appendixC}

In this appendix, by solving Eq.~(\ref{deltak}) in a simple model, 
we show that $\psi_{+s}$, which becomes $e^{ik_{+s}^{(0)}x}$ 
at $x\to\infty$, develops into a superposition of 
two modes given by 
$\sim \exp\left[i{\int^x dx' k_{+s}^{(0)}(x')}\right]$ and by 
$\sim \exp\left[i{\int^x dx' k_{-s}^{(0)}(x')}\right]$ 
in the region of small $x$. 
Here, we assume $\delta k_{+s} \ll 1$ as the condition that 
the expansion with respect to $\delta k_{+s}$ be consistent.

Formally, Eq.(\ref{deltak}) can be integrated easily to obtain 
\begin{equation}
\delta k_{+s}(x) ={i\over 1-v^2(x)} 
e^{-i\int_{x_0}^{x} {2\omega \over 1-v^2(x')}dx'}
\int^{x}_{\infty} \,dx'  
e^{i\int_{x_0}^{x'}{2\omega \over 1-v^2(x'')}dx''} 
 H_+(x'), 
\label{score}
\end{equation}
where we introduced a constant $x_0$, for definiteness, 
although the expression (\ref{score}) is independent of $x_0$.
The integration of $\delta k_{+s}$ becomes
\begin{eqnarray}
\int_{\infty}^x \delta k_{+s}(x') dx'  & = & 
-\int_{\infty}^x dx' \left[\int_{x_0}^{x'}dx'' 
   e^{i\int_{x_0}^{x''} {2\omega \over 1-v^2(x''')}dx'''} 
     H_+(x'')\right] 
{d\over dx'}\left(
{1\over 2\omega}e^{-i\int_{x_0}^{x'} {2\omega \over 1-v^2(x'')}dx''}
\right)
\cr & = &
-{1\over 2\omega}e^{-i\int_{x_0}^x {2\omega \over 1-v^2(x')}dx'}
\int_{\infty}^x dx' 
e^{i\int_{x_0}^{x'}{2\omega \over 1-v^2(x'')}dx''}H_+(x') 
+{1\over 2\omega}\int_{\infty}^x dx' H_+(x'),
\label{delkint}
\end{eqnarray}
where we used an integration by parts for the second equality.
Thus $\psi_{+s}(x)$ is expressed as
\begin{eqnarray}
\psi_{+s}(x) & \approx & e^{i \int^x k_{+s}^{(0)}(x') dx'} 
                    e^{i \int_{\infty}^x \delta k_{+s}(x') dx'}
\cr &\approx &
e^{i \int^{x} k_{+s}^{(0)}(x') dx'}
  \left(1+ i \int_{\infty}^x \delta k_{+s}(x') dx'\right)
\cr & = & 
\left(1+{i\over 2\omega}\int_{\infty}^x dx' H_+(x')\right)
   e^{i\int^x k_{+s}^{(0)}(x') dx'}
\cr &&
-\left({i e^{i\varphi}\over 2\omega}
\int_{\infty}^x dx' e^{i\int_{x_0}^{x'}
   {2\omega \over 1-v^2(x'')}dx''} H_+(x')\right)
 e^{i\int^x k_{-s}^{(0)}(x') dx'},
\label{hosei}
\end{eqnarray}
where $\varphi$ is a real constant. In the last step, Eq.(\ref{delkint}) and 
$-{2\omega\over 1-v^2}=k_{-s}^{(0)}-k_{+s}^{(0)}$ were used.
Let us denote the coefficient of
$\exp\left[i{\int^x dx'k_{-s}^{(0)}(x')}\right]$ on the right
hand side by $\beta$. 
As the probability for the waves to be scattered inward is
proportional to $|\beta|^2$, 
it will be manifest that this scattering probability 
is not generally zero. 

As a simple example, let us consider the case that 
the spacetime is flat, i.e., $\tilde v\equiv 0$ but 
$t$-constant hypersurfaces can fluctuate randomly.
We assume $\gamma_1(x):=\gamma(x)-\gamma_0\ll 1$, 
where $\gamma_0$ is a $x-$independent constant. 
Furthermore, we assume that fluctuations exist 
just in the interval between $x_0-\Delta$ and $x_0+\Delta$. 
We assume that the fluctuations obeys the Gaussian random 
statistics characterized by  
\begin{equation}
 n(\nu):=\int dy\, e^{i\nu y}
  \langle\gamma_1(x) \gamma_1(x+y)\rangle,
\end{equation}
where we used $\langle~\rangle$ to represent the ensemble average. 
Then, the Fourier transformation of $\gamma_1(x)$,
\begin{equation}
 \tilde \gamma_1(\nu):=
  {1\over 2\pi}\int dx\, e^{i\nu x}\gamma_1(x),
\end{equation}
satisfies
\begin{equation}
\langle \tilde\gamma_1(\nu)\tilde\gamma_1^{*}(\nu')\rangle
 ={1\over 2\pi} n(\nu) \delta (\nu-\nu').
\label{corr}
\end{equation}

By setting $\tilde v\equiv 0$,
in Eqs.(\ref{Omegadef}) and (\ref{newv}), 
we find
\begin{equation}
 v(x)=\gamma(x), \quad {1\over \Omega(x)}=\sqrt{1-\gamma^2(x)}.
\end{equation}
Using these equations, we obtain
\begin{equation}
 e^{i\int_{x_0}^x 2\omega\left({1\over 1-\gamma^2(x')}-
 {1\over 1-\gamma_0^2}\right) dx'} H_+(x)
 =H_0+{\omega^4\over k_0^2} \int d\nu\, 
  \left[h(\nu) e^{-i\nu x}-{4\gamma_0 e^{-i\nu x_0}
  \over (1-\gamma_0)(1+\gamma_0)^5}{\omega\over \nu}\right]
 \tilde\gamma_1(\nu) + O(\gamma_1^2), 
\label{a8}
\end{equation}
with
\begin{eqnarray}
h(\nu) = && -{4 \gamma_0 \over (1-\gamma_0)(1+\gamma_0)^{5}} 
{\omega \over \nu}
+{2(-2+\gamma_0)\over (1+\gamma_0)^4}
+{2(3-\gamma_0)\over(1+\gamma)^{3}}{\nu \over \omega}
\cr && +  
{(-8+3\gamma_0)\over 2(1+\gamma_0)^{2}}\left({\nu \over \omega}\right)^{2}  
+{(2-\gamma_0)\over 2(1+\gamma_0)}\left({\nu \over \omega}\right)^{3}.
\end{eqnarray}
The second term in the square bracket in Eq.~(\ref{a8}) does not 
depend on $x$ and the contribution to $\beta$ from this 
term can be neglected. 

Thus, we find that the coefficient $\beta$ evaluated  
in the region $x<x_0-\Delta$ is given by  
\begin{eqnarray}
\beta &:= & -{ie^{i\varphi}\over 2\omega}\int_{\infty}^x dx'\, 
 \theta(x'-(x_0-\Delta))\theta((x_0+\Delta) -x') 
 e^{i\int_{x_0}^{x'} {2\omega\over 1-v^2(x'')}dx''} 
 H_+(x') \cr
&\approx& {i e^{i\varphi}\over 2\omega}
 \int_{x_0-\Delta}^{x_0+\Delta} dx'\,{\omega^4\over k_0^2}
 \int d\nu\, h(\nu)\tilde\gamma_1(\nu) 
 e^{{2i\omega\over 1-\gamma_0^2}(x'-x_0)} e^{-i\nu x'}
\cr &=&
 {i e^{i\varphi}\omega^3\over 2 k_0^2} 
\int d\nu\, h(\nu)\tilde\gamma_1(\nu) 
{2\sin\left((\tilde\omega-\nu)\Delta\right)
         \over (\tilde\omega-\nu)}e^{-i\nu x_0},
\end{eqnarray}
where we introduced $\tilde\omega:=2\omega/(1-\gamma_0^2)$.
Then, with the aid of Eq.~(\ref{corr}), $
\langle|\beta|^2\rangle$ is evaluated as
\begin{equation}
 \langle|\beta|^2\rangle\approx 
 {\omega^6\over 2\pi k_0^4}\int d\nu\, \vert h(\nu)\vert^2 
 \left({\sin\left((\tilde\omega-\nu)\Delta\right)\over (\tilde\omega-\nu)}
 \right)^2 n(\nu).
\end{equation}
If $\Delta$ is sufficiently large,
we can use the approximation
\begin{equation}
  \left({\sin\left((\tilde\omega-\nu)\Delta\right)\over 
  (\tilde\omega-\nu)}\right)^2 \approx \pi \Delta 
    \delta(\nu-\tilde\omega). 
\end{equation}
Therefore, finally we obtain 
\begin{eqnarray}
 \langle|\beta|^2\rangle & \approx &
 {\omega^6 \Delta\over 2 k_0^4}
 \left\vert h\left({2\omega\over 1-\gamma_0^2}\right)\right\vert^2 \,
 n\!\left({2\omega\over 1-\gamma_0^2}\right) 
\cr 
 &\approx& {2 \omega^6 \Delta\over k_0^4}
{\gamma_0^2 \over(1-\gamma_0)^6}
 \, n\! \left({2\omega\over 1-\gamma_0^2}\right). 
\end{eqnarray}
This expression is essentially 
proportional to $(\omega/k_0)^4 \times (\omega\Delta)$.
Since the scattering probability is also proportional to
$\omega\Delta$, 
the effect can be large for large $\Delta$ in principle.
However, in reality this effect is suppressed because of 
the factor $(\omega/k_0)^{4}$.
If $k_0$ is taken to be a Planck scale, 
the factor $(\omega/k_0)^4$ becomes extremely small, 
and then even the waves coming
from the cosmological distance scale 
will not be affected significantly to
induce some observable effects 
unless extraordinary $\omega$ is concerned.

\section{derivation of the conserved current}
\label{appendixD}

Here we derive the conserved current $j$ given in 
Eq.~(\ref{conserved}). 
First we note that ODE~(\ref{ode}) can be derived 
from the variational principle of the action
\begin{equation}
 S_\omega=\int dx {\cal L},
\end{equation}
with
\begin{eqnarray}
 {\cal L}&:=&{1\over 2}\Biggl\{[(-i\omega+v\partial_x)\phi]\cdot
            [(i\omega+v\partial_x)\bar\phi]-
       (\partial_x\phi)(\partial_x\bar\phi)\cr 
&& \quad +
  {1\over 2k_0^2}\left( 
    {1\over \Omega}\partial_x{1\over\Omega}\partial_x\phi
     \right)\partial_x^2 \bar\phi
+
  {1\over 2k_0^2}\left(
    {1\over \Omega}\partial_x{1\over\Omega} \partial_x\bar\phi
     \right)\partial_x^2 \phi\Biggr\}.
\end{eqnarray}
This Lagrangian, ${\cal L}$,   
is invariant under a global phase transformation of $\phi$ given by 
$\phi\to e^{i\lambda}\phi$ and $\bar\phi\to e^{-i\lambda}\bar\phi$.
By using a trivial extension of the standard technique 
to derive the Noether current, we can show that 
\begin{equation}
 j=-i\left\{\left[{\partial{\cal L}\over \partial(\partial_x\phi)}\cdot \phi
 -\left(\partial_x\left({\partial{\cal L}\over \partial(\partial_x^2\phi)}
   \right)\right) \cdot \phi+
 {\partial{\cal L}\over \partial(\partial_x^2\phi)}
  \cdot \partial_x\phi\right] - [\phi\leftrightarrow\bar\phi]\right\},
\label{conserved}
\end{equation}
becomes a conserved current which satisfies $\partial_x j=0$, 
although the present Lagrangian does not have the standard form 
in the sense that it contains $\partial_x^2\phi$.  
Here we adopted the rule that the differentiation 
with respect to $\phi$ or $\bar\phi$ 
is performed as if $\phi$ and $\bar\phi$ are independent.

Applying the formula (\ref{conserved}) to the present case 
with the substitution $\phi=e^{i\int^x k(x) dx}=e^{i\int^x 
 (k_{(R)}+ik_{(I)}) dx}$, we obtain Eq.~(\ref{conserved}).


\begin{thebibliography}{99}
\bibitem{Hawking} S.W.~Hawking, Commun.~Math.~Phys. {\bf 43}, 199 (1975);
Nature {\bf 248}, 30 (1974).
\bibitem{BD} N.D. Birrel and P.C.W. Davies, {\it Quantum fields in curved
  space}, Cambridge University Press, Cambridge, (1982).
\bibitem{Unruh} W.G.~Unruh, Phys.~Rev.~Lett. {\bf 46}, 1351 (1981).
\bibitem{Unruh1} W.G.~Unruh, Phys.~Rev.~D{\bf 51}, 2827 (1995).
\bibitem{JacCor} S.~Corley and T.~Jacobson, Phys.~Rev.~D{\bf 54}, 1568 (1996).
\bibitem{Corley} S.~Corley, Phys.~Rev.~D{\bf 57}, 6280 (1998).
\bibitem{will} e.g., C.M.~Will, 
{\it Theory and experiment in gravitational physics: revised edition},
Cambridge university press (1993).  
\bibitem{phili} G.~Amelino-Camelia, J.~Ellis, N.E.~Mavromatos,
D.V.~Nanopoulos, and S.~Sarkas, Nature {\bf 393}, 763 (1998);
P.~Kaaret, astro-ph/9903464.
\bibitem{Hoo} G.'t Hooft, Nucl.~Phys.~B{\bf 256}, 727 (1985).
\bibitem{BMPS} R.~Brout, S.~Massar, R.~Parentani, and Ph.~Spindel
Phys.~Rev.~D{\bf 52}, 4559 (1995).
\bibitem{CoJ} S.~Corley and T.~Jacobson,  Phys.~Rev.~D{\bf 59}, 
4011 (1999).
\bibitem{Visser} M.~Visser, Class. Qunat. Grav., {\bf 15}, 1767
(1998).  
\bibitem{JacKan} T.~Jacobson and G.~Kang, Class. Quant. Grav., 
{\bf 10}, L201 (1993). 
\bibitem{lattice} S.~Corley and T.~Jacobson, Phys.~Rev.~D{\bf 57}, 
6269 (1998).
\end{thebibliography}
\end{document}